\documentstyle[aps,prb,epsf]{revtex}

\draft

\newcommand \be {\begin{equation}}
\newcommand \ee {\end{equation}}
\newcommand \bea {\begin{eqnarray}}
\newcommand \eea {\end{eqnarray}}
\newcommand \eps {\epsilon}

\newcommand \s {\sigma}

\begin{document}

\title{Inherent structures and non-equilibrium dynamics of
       1D constrained kinetic models: a comparison study}

\author{A. Crisanti$^{1}$\footnote{Andrea.Crisanti@phys.uniroma1.it}, 
        F. Ritort$^{2}$\footnote{ritort@ffn.ub.es},
        A. Rocco$^{1,2}$\footnote{Andrea.Rocco@phys.uniroma1.it}
        and M. Sellitto$^{3}$\footnote{msellitt@ens-lyon.fr}
       } 

\address{$^{1}$ Dipartimento di Fisica, Universit\`a di
Roma ``La Sapienza'', P.le Aldo Moro 2, I-00185 Roma, Italy \\ Istituto
Nazionale Fisica della Materia, Unit\`a di Roma}

\address{$^{2}$ Departaments FFN i ECM, Facultat de F\'{\i}sica, 
           Universitat de Barcelona\\ 
           Avda. Diagonal 647, 08028 Barcelona, Spain}

\address{$^{3}$ Laboratoire de Physique de l'\'Ecole Normale 
             Sup\'erieure de Lyon \\
  46 All\'ee d'Italie, 69007 Lyon, France.}

\author{Version 2.7.3}

\date{September 11, 2000}

\maketitle

\begin{abstract}
We discuss the relevance of the Stillinger and Weber approach to
the glass transition investigating the non-equilibrium 
behavior of models with non-trivial dynamics, but with simple 
equilibrium properties.
We consider a family of 1D constrained kinetic models, which
interpolates between the asymmetric chain introduced by 
J\"ackle and Eisinger [Z. Phys. {\bf B84}, 115 (1991)]
and the symmetric chain introduced by Fredrickson and Andersen
[Phys. Rev. Lett {\bf 53}, 1244 (1984)],
and the 1D version of the Backgammon model 
[Phys. Rev. Lett. {\bf 75}, 1190 (1995)].
We show that the configurational entropy obtained from the inherent 
structures is the same for all models irrespective of their different 
microscopic dynamics.
We present a detailed study of the coarsening behavior of these models,
including the relation between fluctuations and response.
Our results suggest that any approach to the glass transition
inspired  by mean-field ideas and resting on the definition 
of a configurational entropy must rely
on the absence of any growing characteristic coarsening pattern.

\end{abstract} 



\section{Introduction}
There is an old debate concerning the correct description of dynamics
in the glassy state \cite{REVIEW}. According to the general
wisdom, undercooled liquids are in a locally equilibrated metastable
phase, but fall completely out of equilibrium when the relaxation time
exceeds the observation time.  In this situation the glass ages and
slowly relaxes towards equilibrium. While it is widely accepted that
the glass transition observed in laboratory is a purely kinetic
phenomenon, it is still not clear whether a true (or what kind of) 
ergodicity breaking
underlies the glassy behavior and whether the properties
associated to a possible equilibrium transition manifest themselves on
the experimentally accessible time scales.  A particularly interesting 
problem concerns the precise mechanism leading to a slow relaxation 
and its relation with the ground states structure of the system.
 
A possible description of the non-equilibrium regime is in 
terms of coarsening. 
The coarsening process is described by a length
scale which grows in time driving the system towards equilibrium. 
The
most typical scenario for a coarsening dynamics is found in a
ferromagnet quenched down to a temperature below its critical
temperature $T_c$. After quenching, domains of positive and
negative magnetization grow with time. The system acquires a macroscopic 
magnetization only when
the typical domain size becomes of the order of the system size, leading to a
nucleation process which involves overturning of large domains in favor
of the dominant phase. 
Although it cannot be excluded that some type of coarsening behavior similar 
to that found in ferromagnets takes place in real glasses,
there is no strong evidence, up to now,  that any type of coarsening 
process occurs in the relaxation of an undercooled liquid. 

Another possible description for the observed non-equilibrium behavior
calls for activated dynamics.
The activated dynamics scenario is rather different from coarsening. 
No typical growing length scales are now present.  
The system approaches a disordered state, which has no correlation 
with the crystal state, via thermally activated jumps among
different configurations corresponding to structural rearrangements of
spatially localized regions. In this scenario, the ordered crystal state 
has no special relevance. It occurs when fluctuations nucleate a
crystalline droplet of size bigger than a given critical size,
strongly dependent on external parameters such as
temperature. Consequently crystallization can be completely inhibited by
going to low enough temperatures.  It is generally assumed
that crystallization plays a role only for time scales much larger than
those relevant for the relaxation of thermodynamic quantities in the
undercooled phase (such as the enthalpy or specific heat). 
Indeed a glass transition also emerges for purely disordered
systems where a crystal state does not exist \cite{BMe,MPR}. 
This does not exclude the presence of heterogeneities in the glassy phase, 
whose role and influence on the dynamics still needs to be 
understood \cite{HETERO,PARISI}.

Despite the enormous effort devoted to this subject it is still not
clear how these two scenarios combine together into a final
description of the glassy state. Moreover, if coarsening is the driving
process for the relaxation of undercooled liquids, due to the
completely unknown microscopic structure of the
glass state (on which the system should asymptotically relax) it
is unclear what should be experimentally measured in order to identify the
growing domains. 
A similar problem is encountered in spin glasses \cite{BOUCHAUD}. 
Since a disordered state has not a periodic
structure, a priori one does not know how to observe domains and
the question of the nature of the domains in spin glasses is still
unresolved \cite{REVIEW-SG}. However, for spin glasses such
freezing of temporal spin correlations leads to a divergence of the
spin-glass susceptibility. Despite some experimental \cite{NAGEL} and
simulation results \cite{PARISI,SILVIO}, strong evidence for such a
divergence is not found in structural glasses.
The greatest difficulty in elucidating this issue is that a general
non-equilibrium theory to deal with this class of systems is still
missing and approximations to this problem remain partial. They 
usually work
either in a limited range of time scales or in a limited range of
temperatures (for instance, mode-coupling theory \cite{MCT}).

During the eighties a novel approach to the glass transition was
proposed by Stillinger and Weber \cite{SW82}. This 
description of the undercooled liquid phase, inspired
by the Adams-Gibbs-Di Marzio \cite{AGM} theory, incorporates those
features of the energy landscape relevant to the activated regime. 
The Stillinger-Weber approach is based on a decomposition of the 
configurational space into basins (also called inherent structures, 
hereafter referred to as IS) on the basis of the topology of the 
potential energy surface. This construction yields a definition of 
a configurational entropy in terms of dynamically accessible basins 
and it is close in spirit 
to the equilibrium configurational entropy or
complexity of mean-field spin-glass models \cite{KTW,MG}. 

In this paper we study the relevance of the IS analysis proposed
by Stillinger and Weber for non-equilibrium dynamics 
using one-dimensional constrained kinetic models.
Kinetically constrained models were first proposed by Fredrickson
and Andersen \cite{FA} in the attempt to provide a simple microscopic
mechanism for understanding the purely dynamical transition predicted by
the mode-coupling theory. In these models the slowing down
of the dynamics is obtained through the introduction of dynamical constraints,
compatible with detailed balance and Boltzmann-Gibbs distribution.
What makes highly peculiar the relaxational behavior of these models is 
the fact that the slowing down of dynamics is only due to kinetic 
constraints, which prevent certain transitions from occurring.
For a review on early results on these models see Ref. \onlinecite{CONST}.

We find that the non-equilibrium properties of this class of models 
cannot be described in terms of the 
Stillinger and Weber approach, in contrast to what is found in other
models of structural glasses such as Lennard-Jones glasses
\cite{MG,CMPV,SKT99,KST99} or
finite-size fully connected disordered spin glasses
\cite{CR}. 
The main reason is that even though the dynamics is different, 
the configurational
entropy derived from the Stillinger and Weber decomposition is the same for
all models.
While this conclusion is probably valid for all type of coarsening models,
it is an open problem whether the precise mechanism for which the inherent
structure entropy happens to coincide is more general 
and independent of coarsening behavior, i.e. it could also  
hold for higher dimensional constrained models with a different dynamics.

The paper is organized as follows. In Section II we introduce the family of
one-dimensional models studied. In Section III we discuss
the Stillinger and Weber approach and the configurational entropy for
these models. In Section IV we analyze and compare 
the non-equilibrium dynamics for the different models, 
in particular we consider the coarsening phenomena and the 
fluctuations response relation. Finally in Section V we present 
conclusions and discussion. 
Some technical points are analyzed in two Appendices.

\section{The models}

We consider two different classes of 1D constrained models: 
the constrained Ising chain and the Backgammon (BG)
model. For the former we shall mainly consider the
two extreme cases of the symmetrically constrained chain
(SCIC) introduced by Fredrickson and Andersen \cite{FA} and the
asymmetrically constrained Ising chain (ACIC) introduced by J\"ackle and
Eisinger \cite{JE}.  The BG model \cite{BG} is not a purely constrained
kinetic model since there are not local constraint on the microscopic 
dynamics. The constraint here follows from the requirement that the
number of particles remains constant. This induces a global dynamical
constraint which slows down the dynamics as temperature is lowered.
The study of the BG model complement our investigation
comparing its behavior with the locally constrained Ising chain models.

\subsubsection{The constrained Ising chain}

The Hamiltonian of the model is defined by
\begin{equation}
E=-\sum_{i=1}^V\s_i,
\label{eq1}
\end{equation}
where $\s_i$ are Ising-like variables, which take the values $0,1$,
and the index $i$ runs over the sites of a D-dimensional lattice of volume
$V=L^D$. The model corresponds to a paramagnet in a field. 
The dynamics is of the Glauber type where the
spins are randomly updated according to the following rule
\begin{equation}
{\cal W}(\s_i\to 1-\s_i)=
    \left[
    1-\frac{1}{D} \left(\sum_{\mu=1,D}(a\s_{i+\mu}+b\s_{i-\mu}) \right)
    \right] \,\times\,
\min \bigl [ 1,\exp(-\beta\Delta E)\bigr ],
\label{eq2}
\end{equation}
with $a$ and $b=1-a$ positive real numbers. In this paper we consider 
the 1D case $D=1$ although the model is also
interesting for larger dimensions. With exception of the configuration
with all spins equal to $1$ it is known that the space
of configurations is an irreducible Markov chain, so that ergodicity is
guaranteed and detailed balance is fulfilled. 

Depending on the values of
$a$ we may have different cases. In particular for $a=b=\frac{1}{2}$ the model
corresponds to the SCIC \cite{FA},
\begin{equation}
{\cal W}(\s_i\to 1-\s_i)=
            \frac{1}{2}(1-\s_{i+1}+\s_{i-1})
      \,\times\,
     \min \bigl[1,\exp(-\beta\Delta E)\bigr ]
\label{eqSCIC}
\end{equation}
so that a spin can flip iff at least one of its neighbors is down.

If $a$ is equal to $0$ or $1$ then the model corresponds to the 
ACIC ($a=0$) \cite{JE}, 
\begin{equation}
{\cal W}(\s_i\to 1-\s_i)=\min \bigl[ 1,\exp(-\beta\Delta E)\bigr]
             \,\times\, \delta_{\s_{i-1},0}.
\label{eqACIC}
\end{equation}
In this case a spin can flip iff its left neighbor is down, and the
dynamics is more constrained than in the SCIC. 
For generic values of $a$ the
flipping of the spins may take a different probability depending if the
left or right spin is down. 

The vast majority of works appeared in the
literature focused on the previous two cases
(\ref{eqSCIC}) and (\ref{eqACIC}). In the present paper, for completeness 
and for the scope of our discussion, we shall discuss when possible the
behavior for generic $a$.

The dynamics of these models cannot be solved exactly, even if 
several important results are known. 
(i) The correlation time diverges
in the low temperature limit as $\tau\sim\exp(2\beta)$ for the SCIC model 
\cite{RJ,ST1} and as $\tau\sim\exp(\beta^2/\log(2))$ for the ACIC model
\cite{MJ,SE}. 
(ii) In the SCIC model the staggered correlation functions
relax exponentially fast with time and one can define two
characteristic time scales\cite{FR}: the first one 
$\tau=1/(1+\exp(-\beta))$ non-activated, and the second one 
$\tau_1=\exp(\beta)$ activated. In particular the later defines the 
time scale before which no aging effects are seen in the correlation 
functions \cite{FR}. 
(iii) In
the SCIC model the hierarchy of dynamical equations can be
exactly closed for $T=0$ \cite{FR,ST1}.
In the Appendix A we show that this result, 
originally obtained for the SCIC model, can be easily
extended to generic values of $a$. 

\subsubsection{The Backgammon (BG) model}

The energy (Hamiltonian) of the Backgammon model is \cite{BG}
\begin{equation}
E=-\sum_{i=1}^N\delta_{n_i,0},
\label{eq4}
\end{equation}
where $n_i=0,1,\ldots,N$ counts the particles in each site (box) of
a D-dimensional lattice of $N=L^D$ sites.
The energy is given by the number of empty boxes. As before we consider
the 1D case, where $D=1$.
The dynamics is of the Kawasaki type where the particles are randomly
moved from one box to another and the change is accepted with probability
\begin{equation}
 {\cal W} = \min\bigl[ 1, \exp(-\beta\Delta E)\bigr].
\end{equation}
Strictly speaking the BG model is not a constrained kinetic model
since there are not local constraints on particles (or boxes).
Nevertheless the conservation of particles number introduces a global
constraint which makes the dynamics of the BG model
glassy at low temperatures where a lot of particles
accumulate in a small number of boxes and the further emptying of boxes
becomes slower and slower as time goes on. 

In the original definition the
dynamics of this model was mean-field in the sense that particles could move
from one box to any other box. This dynamics can be closed exactly
\cite{BG-closed} and many results have been obtained on its
non-equilibrium behavior \cite{SF,GL}.  Here we are interested in the
equivalent one-dimensional case where boxes are located on a closed ring
(a chain with periodic boundary conditions) and particles can only move from
one box to its left or right neighboring boxes. 
In this case the dynamics at low temperatures
is driven by the coarsening of clusters of empty boxes
similarly to that of spin-1 domains in the kinetically
constrained Ising chain. Here, however, in addition to 
temperature-activated processes induced by energy jumps, the system has
entropic barriers which contribute in a non trivial way to the
coarsening dynamics.  In the original model \cite{BG} with mean-field dynamics
coarsening was absent and
the entropic barriers were the only responsible for glassy dynamics
leading in that case to an activated behavior.

All these models share the common fact that, despite their dynamics,
the thermodynamics is trivial 
and there are no equilibrium spatial
correlations at any temperature. In particular, they do not show any
finite-temperature phase transition. Therefore
the non-trivial behavior follows only from the dynamics, which, 
dynamically constrained
in the first case and ruled by entropic barriers in the second case, 
turns out to be glassy. 

In the next Sections we present a detailed investigation of the
non-equilibrium dynamics of these models and discuss how their 
dynamics cannot be efficiently described in terms of
an IS based configurational entropy approach.

\section{The Stillinger and Weber entropy}

\subsection{The Stillinger and Weber approach}

An interesting approach to investigate activated behavior in glasses
was suggested in the eighties by Stillinger and Weber \cite{SW82}. 
This is based on the (natural) decomposition of the motion near the
glass transition into intra-basin motion (within a
valley) and inter-basin motion (jumps between
valleys). In a ``cage'' picture the first motion 
corresponds to the motion of particles within a cage, while the second 
one to the creation or destruction of cages. This approach
implements in a practical way the old idea
that in the undercooled liquid a strong separation of timescales
of the two motions occurs near the glass transition.
The approach bears some resemblance to the Edwards packing 
entropy in the context of granular systems \cite{EDME,MOPO}. 

Within the Stillinger and Weber (SW) approach each configuration of the
system is mapped into a local minimum of the energy through a local
potential energy minimization which start from the given configuration.
The local minimum was called inherent structure (IS), while the set of
configurations flowing into it defines the basin of attraction or
valley of the IS.
Following SW one then constructs a IS-based thermodynamics decomposing 
the partition function sum into a sum over IS with the same energy 
\cite{SW82}
\begin{equation}
\label{eqZ}
 {\cal Z}(T) \simeq \sum_{e} P_{IS}(e,T),
\end{equation}
with 
\begin{equation}
 P_{IS}(e,T)=\exp\, N\,\left[-\beta e + s_c(e) -\beta f(\beta,e)
\right],\label{eqpe} 
\end{equation} 
where $s_c(e)$, defined as the configurational
entropy,  yields the number of different IS with energy $e$:
 $\Omega(e)=\exp(Ns_c(e))$.
The term $f(\beta,e)$ accounts for the free energy of
the IS-basin of energy $e$, i.e., 
the partition sum restricted to the basin of attraction of IS with energy 
$e$. In each IS-basin the energy has been shifted, so that the IS 
has zero energy, and $f$ accounts only for energy differences.
In general $f(\beta,e)$ 
may have a non-trivial dependence on the
energy if the IS-basin of IS with different energy are quite different. 
When the temperature is such that only the states near the bottom of the
IS-basin do contribute then it is reasonable to expect that $f(\beta,e)$ 
is roughly independent of $e$ \cite{SKT99,KST99,CR}.
Another case where the dependence of
$f(\beta,e)$ on $e$ is expected to be negligible is when the 
IS-basins are narrow and contain few configurations.
This approximation works very well for REM-like models \cite{DERRIDA,CR}.
When the $e$-dependence of $f$ can be neglected the configurational
entropy $s_c(e)$ can be obtained directly from (\ref{eqpe}).

As long as the configurations counted on the r.h.s. of eq. (\ref{eqZ})
are the most relevant for the thermodynamics at temperature $T$ the
above construction is totally legitimate as far as thermodynamics is
concerned. It is only a different way of summing the partition function.
Nevertheless the $s_c(e)$ obtained with the SW decomposition 
is in some sense a dynamical quantity since the projection between 
configurations and IS basins can be seen as the zero-temperature 
dynamics of the model. For this reason we will refer to it as 
Stillinger-Weber configurational entropy to distinguish it from other 
possible definitions of the configurational entropy taken from mean-field 
concepts \cite{MG,CMPV}. 
This poses the question, first raised by Monasson and Biroli\cite{BM},
on the relevance of $s_c(e)$ and IS in general for dynamics.

It is clear that once the energy and the rules of the
dynamics are given the IS can always be defined. For example 
for spin-glass models with quenched disordered variables
taken from a continuous distribution, the dynamics usually consists of
Monte Carlo updates (for instance, single spin-flips). The IS are
then identified as the final configurations reached after a sequence of
Monte Carlo moves where the spin which yields the largest decrease of
energy is identified and flipped. Consequently, IS are stable against
single spin-flips but not for higher-number of spin flips.
Biroli and Monasson \cite{BM} conclude then that IS are ill-defined 
because $s_c(e)$ depends on the number of spin-flips, which make the 
configuration stable (at least, for not-fully connected models). 
We disagree with this conclusions because, as noted above, the 
IS and the SW entropy are intimately related to dynamics, 
and therefore it is not a surprise
that changing the dynamics the IS and the SW entropy in general 
change. What, in our opinion, is ill-defined is
to speak of IS without specifying the dynamics. 

Nevertheless the question posed by Biroli and Monasson is far from 
being trivial. Indeed despite the fact that the IS and the SW are
dynamical quantities it is far from obvious that they contain all
(or almost all) relevant informations on the dynamics on long time scales.
This is a well known problem in the theory of dynamical systems. 
The SW decomposition can be seen as a mapping of the true dynamics
at a given temperature onto a symbolic dynamics given by the dynamics 
of the IS. The obtained symbolic dynamics gives a good description
of the original one only if the mapping defines what is called
a ``generating partition'', see e.g. \cite{BS}. 
In general for a generic dynamics it is not 
at all trivial to demonstrate that such a partition exists, and even 
if it does exist, how to find it.  
We can then recast the question posed by Biroli and Monasson in the following
way: does the SW decomposition lead to a generating partition, or at least to
a good approximation of it, for the long time dynamics of glasses near the
glass transition?

In general we can answer to this question only a posteriori.
We define a possible partition and then check if this reproduces
the desired features of the dynamics.
However we can try to
find under which conditions the answer could be affirmative.
Usually to find a generating partition, or a good approximation
of it, a good starting point is by looking at the ``physical'' 
properties of the dynamics.
The SW mapping replaces each configuration in a IS-basin with the IS
itself. Therefore it is clear that this mapping will be a good mapping 
if the systems spends a lot of time inside the basin. Under this assumption 
the dynamics on time scales larger than the typical residence time inside 
a IS-basin should be quite well described by the IS dynamics. 
This scenario is typical of a many valley dynamics with activated dynamics.
It is also clear from the above discussion that if the IS mapping is a good 
mapping it does not matter which configuration inside the IS-basin
is used to represent the IS-basin. It may be the IS itself or any other 
configuration in the basin. In a recent study of finite-size mean field
spin-glass models, which share the properties of structural glasses, 
this independence has been indeed observed \cite{CR}. 

On the contrary for dynamical processes described by a coarsening
process this description should in general fail because dynamics 
proceeds through geometrically correlated configurations. 
Barrat, Burioni and Mezard have shown \cite{BBM}
that the difference between the two scenarios has a simple
manifestation in how dynamical trajectories departing
from the same configurations  separate in the phase space. 
Consider a system described by a vector $\vec{X}(t)$ in configuration
space. At time $t_w$ the system
is cloned into a new system described by the vector $\vec{Y}(t)$.
The two copies are let evolve with different realizations of thermal noise
and the overlap $Q_{t_w}(t)=\vec{X}(t_w+t)\cdot\vec{Y}(t_w+t)$
is recorded as a function of time $t$. 
For coarsening-like systems (called type I systems in \cite{BBM}) the overlap
converges to a finite value $Q_{\infty}=\lim_{t\to\infty}Q_{t_w}(t)$ for
any value of $t_w$ while for glassy systems with 
structural glasses behavior 
(called type II systems in \cite{BBM}) that limit gives,
for all possible values of $t_w$, the lowest possible value of
$\vec{X}(t_w+t)\cdot\vec{Y}(t_w+t)$. Since in general it is possible to 
define IS also for coarsening systems, this is an indication that in these
systems the non-equilibrium dynamics goes through configurations that 
are unrelated with the IS. Most probably the relevant configurations
for coarsening are those on the border among IS-basins, i.e., those
configurations which are not mapped into any IS.
In this case the IS and the SW configurational entropy can still be defined
but are obviously of little use for understanding non-equilibrium dynamics, 
as shown by the results reported in the following Sections.

Before addressing the SW approach to non-equilibrium dynamics of 
constrained kinetic models we note that 
the definition of the SW configurational entropy may not be completely
free of ambiguities, especially for systems with discrete states.
Indeed the SW mapping assumes that once the energy and the dynamics
are given then  the mapping between configurations and IS is
uniquely defined. This means that regardless of when a given configuration
appears in the dynamical evolution it will always be mapped to the same IS.
In most of the recent papers on this subject \cite{SKT99,KST99,CR}
this was the case, but for discrete models with discretized values 
for the energies, as the ones studied here, 
there may be problems because there could be many 
directions in phase space where the energy decreases by the same amount. 
In this case some ``decision'' must be taken, e.g., one could employ a random
choice among the possible directions.
The IS and the corresponding SW configurational
entropy can still be defined, but now they depend on the chosen strategy
for dealing with equivalent directions. It can be shown \cite{CRu}
that this leads to a temperature dependence of $s_c$, so that
the form of $s_c$ depends on $T$.
Only when some additional requirements are fulfilled the temperature
dependence disappears. For example when all possible strategies lead
to the same sampling rate for the IS relevant for the dynamics.
We anticipate that this requirement is not fulfilled by the constrained
kinetic models nor by the BG model studied here, and in some cases
we find different curves for $s_c$ for different temperatures.
Nevertheless the violation is not too strong since we basically 
find only two different curves depending on the temperature range.
This reflects the trivial fact that the constraint is more or less
effective depending on the temperature because more we lower the temperature
more the system orders. 
Another interesting feature of these models is that, despite the
fact that the properties of these
dynamical models are rather different, 
all of them have the same SW configurational entropy $s_c$. 
This casts doubts on the relevance of the IS analysis for the 
non-equilibrium dynamics of these models.

\subsection{The kinetically constrained Ising chain}

These models are defined by eqs. (\ref{eq1}) and (\ref{eq2}).  
To compute $s_c(e)$ we thermalize the system at a finite temperature $T$. 
This can be achieved either by running the dynamics for a sufficient 
long time or by starting from equilibrium configurations
whose distribution is given by
\begin{equation}
P_{eq}(\s_i)=\frac{\exp(\beta\s_i)}{1+\exp(\beta)}.
\label{eqs1}
\end{equation}
For each thermalized configuration the corresponding IS is
computed via the minimization process given by the
zero-temperature dynamics. Repeating the procedure for several initial
configurations the IS probability distribution (\ref{eqpe}) can be 
evaluated.

Each IS is a fix-point of the dynamics, and therefore we can 
estimate $s_c$ from the number of fix-points.
This is easily evaluated 
denoting with $A$ the one-bit sequence $1$ and $B$ the
two-bits sequence $01$ since all fix-points are given by all 
possible arbitrary sequences of $A$'s and $B$'s (for instance, the sequence
$ABBBAABB$). If
$N_A,N_B$ stand for the number of $A$'s and $B$'s in the sequence, 
with $N_A+2 N_B=N$, where $N$ is the length of the chain, then 
\begin{equation}
N_{fix}=\frac{(N_A+N_B)!}{N_A! N_B!}.
\label{sc}
\end{equation}
 From this expression, and  noting that $e=-(N_A+N_B)/N$,
we have
\begin{equation}
   s_c(e)= \frac{\log(N_{fix})}{N}
         =-e\log(-e) -(1+e)\log(1+e) +(1+2e)\log(-1-2e).
\label{eqs2}
\end{equation}
The above
formula assumes that all IS are counted with an equal {\it a priori}
probability. This is what is called the method of {\it unbiased guess} in
information theory \cite{BS}. Therefore
the above expression is valid iff the dynamics samples (almost) all fix-points 
with equal probability, where with ``almost'' we mean the fix-points
relevant for equilibrium dynamics at temperature $T$. 

A better estimate of $s_c$ comes from the analysis of the zero-temperature 
dynamics. As shown in Appendix A the zero-temperature dynamics 
can be solved exactly for any value of $a$, and from this 
$P_{IS}(e,T)$ can be evaluated:
\begin{equation}
P_{IS}(e,T)=\frac{1}{\sqrt{2\pi\langle C_0^2(\infty)\rangle_c}}\exp\Bigl
( -\frac{(e-\langle e_{IS}\rangle)^2}{2\langle
C_0^2(\infty)\rangle_c}\Bigr).
\label{eqBPDE}
\end{equation}
The details of the calculation are reported in the Appendices
together with the expressions of $\langle C_0^2(\infty)\rangle_c$
and $\langle e_{IS}\rangle$, eqs. (\ref{final}) and (\ref{av_e}).
Note that this result, as well as eq. (\ref{eqs2}), 
does not depend on the value of $a$ implying that all models, and 
in particular the SCIC and ACIC, have the same $P_{IS}(e,T)$ and
hence the same SW configurational entropy.

\begin{figure}
{\epsfxsize=9cm\epsffile{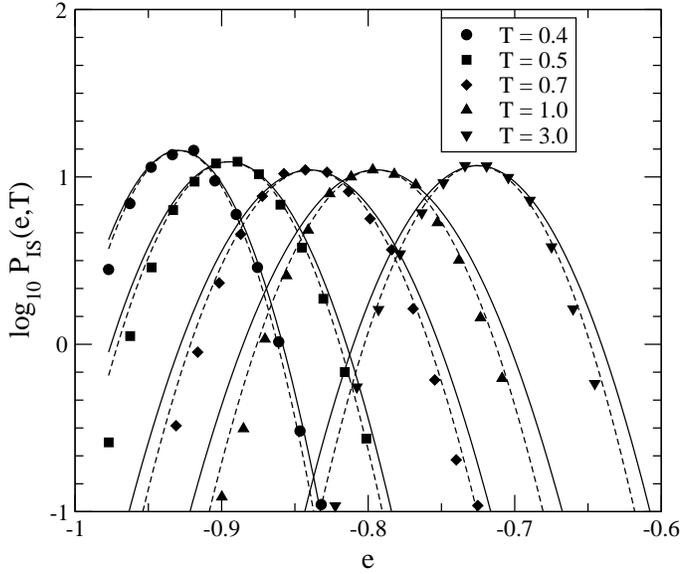}}
\caption{Probability histograms in the SCIC model with 64 spins at
different temperatures compared with the analytical prediction
(\ref{eqBPDE}) (full curve) and a fit to a Gaussian (dashed curve).}
\label{pde0_FA}
\end{figure}

In Fig. \ref{pde0_FA} we compare the numerically evaluated $P_{IS}(e,T)$
for the SCIC model with $N=64$ and different temperatures
with the analytical prediction eq.(\ref{eqBPDE}). The agreement is
quite good at low temperature but decreases with increasing 
temperatures where 
the variance of the Gaussian is slightly larger than in simulations. 

\begin{figure}
{\epsfxsize=9cm\epsffile{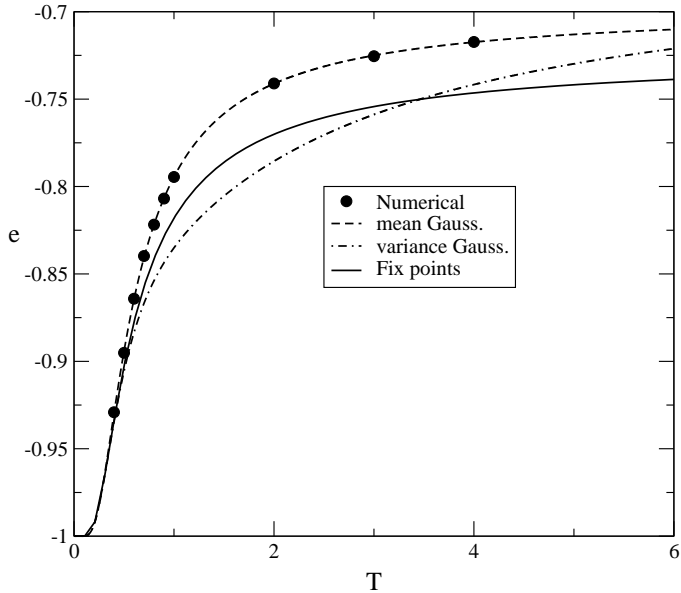}}
\caption{IS energies as function of temperature obtained 
integrating (\ref{eqs2}), using (\ref{av_e}) and the variance of the
IS-energy distribution given by (\ref{eqe2}).}
\label{FAec0}
\end{figure}

In Fig. \ref{FAec0} we report the average IS energy 
as a function of $T$ and compare it with the prediction from the
analytical calculation of Appendix B. There are two possible 
ways to compute $e(T)$: the first is from eq. (\ref{av_e}), the second
rests on integrating the variance
\begin{equation}
e(T) = \int_0^T\frac{\langle C_0^2(\infty)\rangle_c}{T^2}\,dT,
\label{eqe2}
\end{equation}
where $\langle C_0^2(\infty)\rangle_c$ is given by 
(\ref{final}). We also report the result from 
the fix-point approximation (\ref{eqs2}). Fig. \ref{FAec0} clearly 
shows that the different approximations depart each other at 
a temperature $T\simeq 0.6$. Above this temperature the energy dependence of 
the IS free energy $f$ in eq. (\ref{eqpe}) cannot be neglected anymore,
showing that the fix-point approximation which neglects
thermal fluctuations inside the IS-basins
is inappropriate. On the other hand the direct calculation from the 
zero-temperature dynamics turns out to be very good.

\begin{figure}
{\epsfxsize=9cm\epsffile{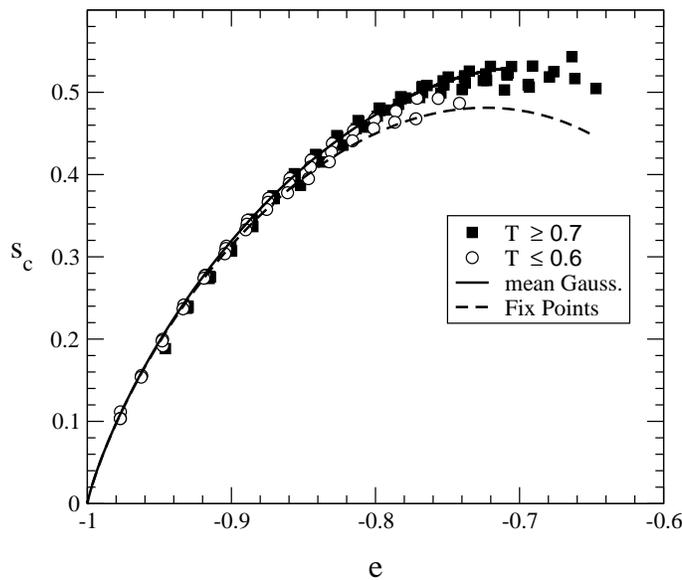}}
\caption{SW configurational entropy in the SCIC for $N=64$ spins at
different temperatures compared with the analytical prediction
(\ref{sc1}) (upper curve) and the fix-point estimate (\ref{eqs2})
(lower curve).}
\label{sc_FA}
\end{figure}

Finally we consider $s_c$. 
In Fig. (\ref{sc_FA}) we show the
results obtained for the SCIC model with $N=64$ and different temperatures. 
The SW configurational entropy $s_c$ is obtained from the numerical 
$P_{IS}(T,e)$ as \cite{KST99,CR}:
\begin{equation}
s_c(e)=\beta e+\log(P_{IS}(T,e)/N) + {\rm const}.
\label{formula_sc}
\end{equation}
For each temperature the constant has been fixed by  
collapsing different data onto the single curve. 
As a comparison we also show the theoretical
predictions from equations (\ref{eqs2}) and [See Appendix B]
\begin{equation}
s_c(e)=\int_{0}^T\frac{d\langle e_{IS}\rangle}{dT}\frac{dT}{T}.
\label{sc1}
\end{equation}
As shown in the Appendix B, both coincide asymptotically
close to the ground state energy $e=-1$. The collapse is excellent
showing that the approximation (\ref{eqs2}) and the low-temperature
behavior (\ref{sc1}) asymptotically coincide in the limit $T\to 0$. 
We note that there is a range 
of energies where data from $T\le 0.6$ collapse on one curve while
data for higher temperature collapse on a different curve. As discussed
above, this 
residual temperature dependence follows from the presence of
many equivalent directions for energy minimization.

We have checked that $P_{IS}(T,e)$ is independent of $a$ by
repeating the analysis for the ACIC model and for different values of $a$.
In all cases we always find the same results.

\subsection{The BG model}

In this case we cannot exactly solve the
zero-temperature dynamics of the model and compute the SW
configurational entropy. 
Nevertheless, we can approximate $s_c(e)$ of the BG model by
counting the number of ways in which two or more particles can be
distributed in a set of $N$ occupied boxes separated by empty boxes. 
This yields
two contributions: the first comes from all possible ways of
distributing the $M$ occupied boxes in a chain of $N$ boxes, with the
additional condition that each occupied box is surrounded by an empty
box. This is again given by eq. (\ref{sc}) assuming that 
$\sigma=1$ for occupied boxes and $\sigma=0$ for empty ones. The
energy (\ref{eq4}) is given by $E=-(N-M)$ and therefore
this contribution reads,
\begin{equation}
s_c^{first}(e)=-(1+e)\log(1+e)-e\log(-e)+(1+2e)\log(-1-2e).
\label{scfirst}
\end{equation}

The second contribution follows from considering all different ways of
distributing the $N$ particles among the $M$ occupied boxes with the
constraint that each occupied box contains at least two particles:
\begin{equation}
N_{fix}^{second}=\sum_{\prod_{r=1}^N n_r\ge 2}\frac{1}{\prod_r n_r!}\delta(\sum_{r=1}^N{n_r}-N).
\label{eqs3}
\end{equation}

where the $n_r!$ terms arise from the distinguishability of particles.
Introducing the integral representation for the delta function,
\begin{equation}
\delta(x)=\frac{1}{2\pi}\int_{-\infty}^{\infty}\exp(i\lambda x)\,d\lambda,
\label{integral}
\end{equation}
we find an expression for (\ref{eqs3}) in terms of the fugacity $y$. In the 
$N\to\infty$ limit this can be
evaluated by the saddle point method yielding
\begin{equation}
s_c^{second}(e)=\frac{\log(N_{fix}^{second})}{N}=-\log(y)+(1+e)
\log(\exp(y)-y-1), 
\label{eqs4}
\end{equation}
where $y$ satisfies the saddle-point condition,
\begin{equation}
e=-1+\frac{\exp(y)-1-y}{y(\exp(y)-1)}.
\label{saddle}
\end{equation}
 
The full entropy is given by
\begin{equation}
s_c(e)=s_c^{first}(e)+s_c^{second}(e).
\label{total}
\end{equation}
Note that for this model the configurational contribution may be
negative because particles are distinguishable. 

\begin{figure}
{\epsfxsize=9cm\epsffile{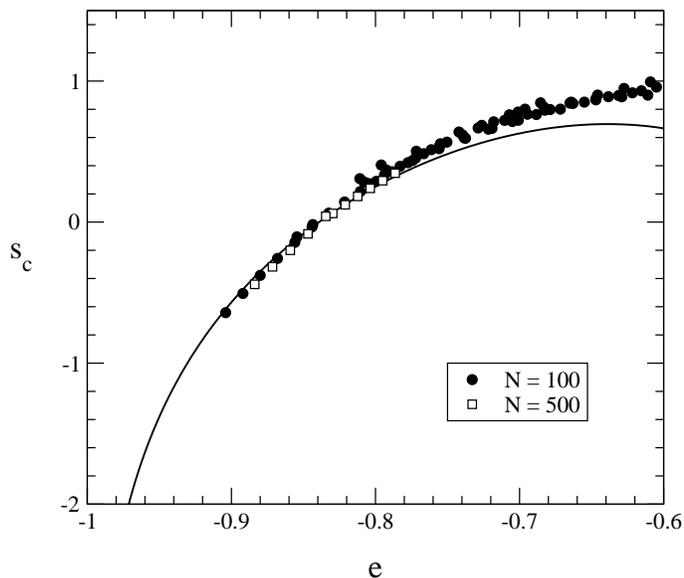}}
\caption{SW configurational entropy in the BG model for $N=100,500$
boxes at different temperatures $T=1.0,0.5,0.4,0.3,0.2,0.15,0.1$ 
compared with the fix-point estimate (\ref{total}) (full line).}
\label{sc_BG}
\end{figure}

We have computed $s_c(e)$ numerically following a procedure similar to that
described for the constrained kinetic models.
The results are shown in Fig. \ref{sc_BG} for two different sizes $N=100,500$
and temperatures ranging from $T=0.1$ up to $T=1$. Similarly to what found for
the constrained kinetic models, the data nicely collapse onto a
single curve although it does not exactly coincide with the number of fix
points. In this model the presence of different equivalent directions to
decrease the energy does not influence $s_c$. This is most probably due to the
global character of the constraint.

Comparing Figs. \ref{sc_FA} and \ref{sc_BG} we see that
the agreement now is worse. We attribute this to the
presence of entropic barriers which follows from all
possible arrangements of particles inside the
boxes. All arrangements leave the energy unchanged, but their
number strongly depends on the number of empty boxes, leading
to a stronger energy dependence of the IS free energy for this model.
This effect was not present in the kinetically constrained Ising chain.

The conclusion that can be drawn from this Section is that for
these models a description of
the glassy behavior in terms of a complex energy landscape is not 
relevant. Even though the SW configurational entropy for the constrained 
Ising chain is a non trivial quantity, it does not distinguish the SCIC model
from the ACIC model.

\section{Analysis of coarsening behavior}

In the previous Section we have seen that the IS approach 
yields identical results for models which are known to have a completely 
different dynamical behavior, namely the SCIC and ACIC models. 
The purpose of this Section is to evidentiate these
difference making connection with results already known in the
literature and studying new ones to gain some insight using the tools
from disordered systems.  Coarsening appears when domains of a
given phase grow in time slowly enough for the system to be
off-equilibrium \cite{BRAY}. In the simplest case, dynamics is
characterized by a unique length scale associated with the typical size of
the growing domains. All models discussed in this paper can be described
in terms of coarsening in the sense that it is possible to
define a length scale which identifies the distance from to equilibrium. 
For the kinetically
constrained Ising chain this length is the typical size of the $+1$
domain while  in the BG model it is the typical length of sequences of empty
boxes. 

In the simplest cases this length suffices to characterize the
off-equilibrium behavior. For instance, for coarsening in
(ordered or disordered) ferromagnets the off-equilibrium behavior is fully
characterized in terms of a single length scale $L(t)$ in the sense that the
two-times dependence of correlation and response functions directly
enters through the value of this length scale. In the aging regime,
where both times $t,s$ are large we have \cite{BRAY}
\begin{equation}
C_{ag}(t,s)\sim f\left ( \frac{L(t)}{L(s)} \right ),
\label{eqC}
\end{equation}
\begin{equation}
R_{ag}(t,s)\sim \frac{1}{L(t)^{\alpha}}\, g\left( \frac{L(t)}{L(s)}
\right ), ~~~~~~~~t>s,
\label{eqR}
\end{equation}
with $\alpha\ge 1$ a positive exponent which depends on the model
under consideration.  

Correlations are easy to measure, while responses require the 
introduction of an
external perturbation. This perturbation must couple with the variables
of the system and must be small enough to ensure a 
linear response regime.  A useful quantity 
is the integrated response function (hereafter denoted as IRF), which
measures how much the system remembers the effects of a perturbation
applied during a given interval of time \cite{REVIEW-DYN},
\begin{equation}
\chi(t,s)=\int_s^t\,du\, R(t,u)
\label{eqI}
\end{equation}
One of the salient features of coarsening phenomena \cite{BRAY,BBK,MK}
is that aging
effects in the integrated response function asymptotically vanish when
the lower time $s$ goes to infinity and the system does not have
long-term memory. Suppose that the coarsening length $L(t)$
grows like $t^{1/z}$ with $z$ a dynamical exponent. Using
eqs.(\ref{eqR}) and (\ref{eqI}) one finds that the aging part of the
IRF behaves like,
\be
\chi_{ag}(t,s)\sim s^{1-\frac{\alpha}{z}}\hat{\chi} (t/s)
\label{eqI_aging}
\ee
which vanishes as $s\to\infty$ if $\alpha>z$. 

An easy way to test these effects is by directly
looking at fluctuation-dissipation plots (hereafter referred to as FDT)
\cite{FRI}.  In equilibrium $R(t,u)=R(t-u)=\beta \frac{C(t-u)}{\partial
u}$ which substituted into (\ref{eqI}) yields,
\begin{equation}
\chi(t-s)=\beta[C(0)-C(t-s)]
\label{eqI2}
\end{equation}
and the plot of $T \chi(t,s)$ in terms of $C(t,s)$ is a straight line of
slope $-1$. 

In the off-equilibrium regime expression (\ref{eqI2})
can be generalized  by defining the
fluctuation-dissipation ratio \cite{REVIEW-DYN}
\begin{equation}
X(t,s)=\frac{TR(t,s)}{\frac{\partial C(t,s)}{\partial s}}   
\label{fdr}
\end{equation}
which measures how far the system is from equilibrium.
In equilibrium $X$ is equal to $1$. In the off-equilibrium asymptotic 
long-time regime, i.e. in the aging regime,  where there is no 
time translational invariance both
$X(t,s)$ and $\chi(t,s)$ are expected to be non-trivial functions of $C$.
A quantitative estimate
of $X$, can be obtained from the slope of the FDT plots: 
\begin{equation}
X(C)=-\left[\frac{\partial\, T\chi(t,s)}
            {\partial C(t,s)}\right]_{C(t,s)=C}~~~~~.
\label{slope}
\end{equation}

For coarsening models the aging part of
the IRF asymptotically vanishes and $\chi(t,s)$ is expected to 
saturate to a finite value, called the field-cooled value in the 
context of spin glasses, and stays constant while the correlation
still decreases before saturating.

The simplest way to compute the IRF in the present class of models is to
apply a perturbation which does not couple with the absorbing state,
i.e., the ground state.  We start from a random initial
configuration and after a waiting time $s=t_w$ apply a random staggered field
$h_i=\epsilon_i\,h_0(t)$, where $h_0(t)$ is the intensity of the field and
$\epsilon_i=\pm 1$ are independent random quenched variables of zero mean.
This method 
has the advantage that the perturbation term in the
Hamiltonian does not directly couple to the coarsening length, and
has been used
by Barrat \cite{BARRAT} to investigate coarsening in finite-dimensional
Ising models.

The staggered magnetic field couples to the spin variables 
$\sigma_i$ in kinetically constrained models 
and to the equivalent variables $\sigma_i=\delta_{n_i,0}$
in the BG model so that the perturbation in the Hamiltonian reads
\begin{equation}
\delta {\cal H}(t)=-\sum_{i=1}^N\,\epsilon_i\,h_0(t)\,\sigma_i~~~~~.
\label{eqP}
\end{equation}

In the case of $h_0(t)=h_0\,\theta(t-t_w)$  the integrated response
function can be obtained measuring the 
random staggered magnetization after switching the field at time $s=t_w$
as,
\begin{equation}
\chi(t+t_w,t_w)=\frac{1}{N\,h_0}\sum_{i=1}^N\eps_i\,\s_i(t+t_w).
\label{eqM}
\end{equation}
The original $0,1$ variables have some disadvantages, for example
the correlation at equal time is not $1$ but depends on temperature.
For this reason we find more convenient to work with 
the new variables $\nu=2\s-1$ which now take the values $1,-1$. 
We then consider the disconnected correlation,
\begin{equation}
C(t+t_w,t_w)=\frac{1}{N}\sum_{i=1}^N\nu_i(t_w)\,\nu_i(t+t_w)
\label{disc_c}
\end{equation}
and the staggered magnetization (\ref{eqM}),
\begin{equation}
M_{stag}(t+t_w,t_w)=\frac{1}{N}\sum_{i=1}^N\epsilon_i\,\nu_i(t+t_w)
                   = 2\, h_0\,\chi(t+t_w,t_w)
\label{disc_m}
\end{equation}
Now the equal times disconnected
correlation function is equal to $1$ so that in the FDT plots,
 where the integrated response function is plotted versus the 
disconnected correlation function, all curves 
start  from $C=1,M_{stag}\simeq 0$ for $t=t_w$. This
makes easier to compare results from different values of
$t_w$.

As discussed before in Section III a different way to distinguish 
coarsening dynamics from other
more complex behaviors is to measure the overlap $Q(t)$ between two
replicas which start from the same configuration at time $t_w$ and 
evolve with different realization of thermal noises. 
Here we consider the connected, normalized 
overlap function,
\be
Q^c_{t_w}(t)=\frac{\frac{1}{N}\sum_{i=1}^N
\sigma^{(1)}_i(t_w+t)\,\sigma^{(2)}_i(t_w+t)
-\bigl(\frac{1}{N}\sum_{i=1}^N\sigma^{(1)}_i(t_w+t)\bigr)\bigl(
\frac{1}{N}\sum_{i=1}^N\sigma^{(2)}_i(t_w+t)\bigr)}{\frac{1}{N}\sum_{i=1}^N
\sigma^{(1)}_i(t_w)-[\frac{1}{N}\sum_{i=1}^N\sigma^{(1)}_i(t_w)]^2}
\label{Qcon}
\ee
where $\sigma^{(1,2)}=0,1$ refer to the two replicas, and 
the connected, normalized, 
correlation
\be
C_c(t+t_w,t_w)=\frac{\frac{1}{N}\sum_{i=1}^N
\sigma^{(1)}_i(t_w+t)\,\sigma^{(1)}_i(t_w)
-\bigl(\frac{1}{N}\sum_{i=1}^N\sigma^{(1)}_i(t_w+t)\bigr)\bigl(
\frac{1}{N}\sum_{i=1}^N\sigma^{(1)}_i(t_w+t)\bigr)}{\frac{1}{N}\sum_{i=1}^N
\sigma^{(1)}_i(t_w)-(\frac{1}{N}\sum_{i=1}^N\sigma^{(1)}_i(t_w))^2}
\label{Ccon}
\ee

For type I systems $Q^c_{\infty}=\lim_{t\to\infty}Q^c_{t_w}(t)$ is finite.
For type II systems this quantity converges in the $t\to\infty$ limit to
the lowest accessible value, i.e. vanishes in the $t\to\infty$ limit. 
In equilibrium both (\ref{Qcon}) and (\ref{Ccon}) depend only on $t$ 
and the following relation is valid $Q^c_{t_w}(t)=C_c(2t)$.
There are few numerical studies of $Q^c_{t_w}$ 
\cite{BBM,AFR}. 
For the models considered here the results from this analysis are,
however,
not so strong, the reason probably being that coarsening occurs in 
a disordered (i.e. paramagnetic) phase.

In the rest of this Section we investigate in detail coarsening length scales,
correlations and  responses for the non-equilibrium dynamics of 
models described in Section II. 
We note that although many results on timescales and coarsening length
scales have been obtained in the literature \cite{JE,RJ,ST1,MJ,SE,FR}
almost nothing is known about the aging behavior in this type of models
(partial results are shown in \cite{FR} for the SCIC).

\subsection{The SCIC model}

\begin{figure}
{\epsfxsize=9cm\epsffile{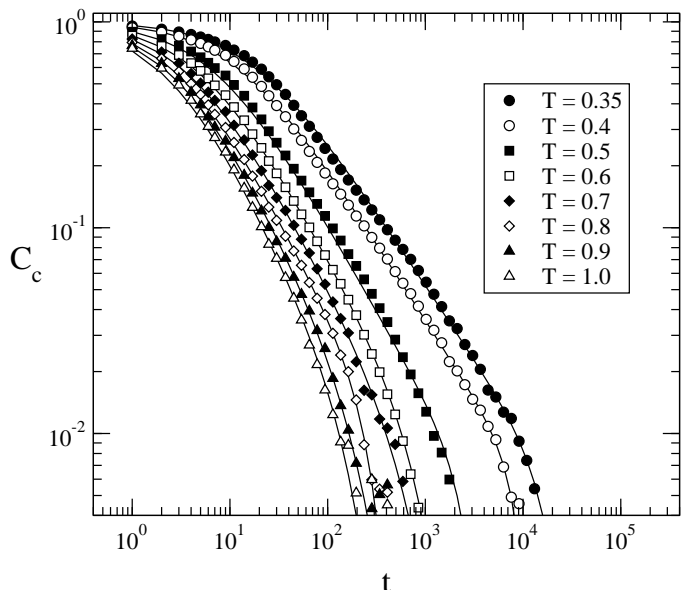}}
\caption{Equilibrium connected correlation functions in the SCIC for 
temperatures
ranging from $T=1.0$ down to $T=0.35$ fitted according to
(\ref{eqC2}). Data have been averaged over 1000 thermalized initial
conditions and $N=10^5$.}
\label{C_FA}
\end{figure}

It has been shown in \cite{FR} that the SCIC model has an activated
timescale $\tau_1=\exp(\beta)$ characterized by an exponential decay of the
staggered energy. For times smaller than $\tau_1$
there are no aging effects and only for times larger than $\tau_1$
non-equilibrium behavior with non-exponential relaxation and aging appear. 
 From the decay of correlation functions a second
activated timescale $\tau_{\rm corr}>\tau_1$ can be defined.
To this end we have computed the equilibrium 
connected correlation function
\be
C_c(t)=\frac{N\sum_{i=1}^N\s_i(0)\,\s_i(t)-\sum_{i=1}^N\s_i(0)\sum_{i=1}^N
\s_i(t)}{N\sum_{i=1}^N\s_i(0)-(\sum_{i=1}^N\s_i(0))^2}
\label{eqC1}
\ee
which is well described for
all temperatures by the following functional form
\be
C_c(t)=\left (\frac{1}{1+ t/\tau_1}\right )^{\alpha}\exp(-at^b),
\label{eqC2}
\ee
where $\alpha,a,b$ are three fit parameters and the activated 
timescale $\tau_1$
is introduced in the fitting function as an effective microscopic time. 
The results for $C_c(t)$ are shown in Fig. \ref{C_FA} for different
temperatures,
the lines are the best fits with form (\ref{eqC2}). This fits are
in agreement with the asymptotic analytical predictions of Reiter and
J\"ackle \cite{RJ} and Schulz and Trimper \cite{ST1} but combined with
the exponential timescale $\tau_1$ derived in \cite{FR}. In particular
the exponent $\alpha$ is close to the value
$1/2$ predicted in \cite{RJ,ST1} for very low temperatures. 

\begin{figure}
{\epsfxsize=9cm\epsffile{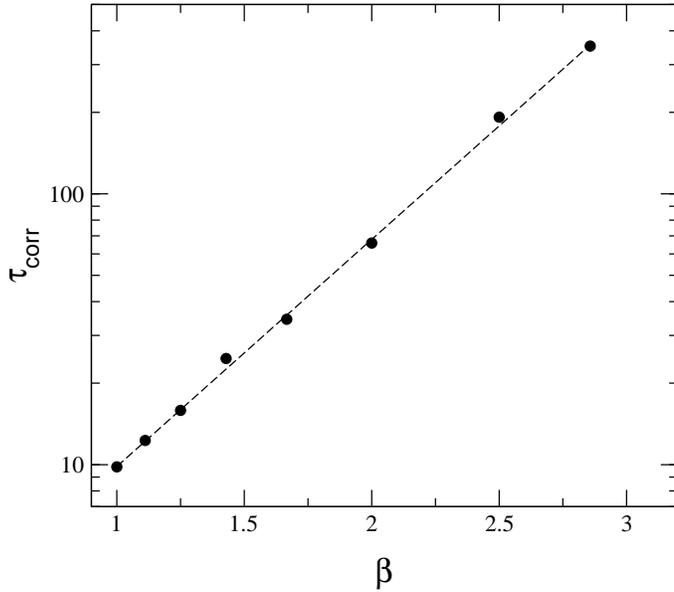}}
\caption{Correlation time in the SCIC computed using (\ref{int}) 
         and fitted with an Arrhenius behavior. The best fit gives
         $\tau_{\rm corr}=1.43\exp(1.93\beta)$}
\label{decorr}
\end{figure}

 From the fit we can estimate $\tau_{\rm corr}$ as
\be
\tau_{\rm corr}=\int_0^{\infty}C_c(t)\,dt~~~~~.
\label{int}
\ee
The results are shown in Fig. \ref{decorr}. The
correlation time $\tau_{\rm corr}$ follows the Arrhenius type law
$\tau_{\rm corr}\sim \exp(2\beta)$. 
This functional dependence of 
correlation time from temperature can be understood from
the following phenomenological argument, based on
defects annihilation in the SCIC. A defect
separated by magnetized domains can disappear by
anchoring defects along the chain. The typical time to anchor a defect
is $\exp(\beta)$ while  the length of the magnetized domains in equilibrium
is of order $\exp(\beta)$. Because defects can be anchored starting
form the right or from the left of a magnetized domain, the typical
time to annihilate that domain is the sum of independent processes
yielding  $\tau_{\rm corr}\sim \exp(2\beta)$. 

To investigate coarsening in the SCIC model we have measured the growth
of the average domain length
\be
d(t)=\frac{\sum_{l=1}^N\,l\,P_t(l)}{\sum_{l=1}^N\,P_t(l)}
\label{d}
\ee
where
\be
P_t(l)=\sum_{i=1}^N\prod_{j=i}^{i+l-1}\sigma_j(t)\,[1-\sigma_{i+l}(t)][1-\sigma_{i-1}(t)]
\label{Pl}
\ee
is proportional to the probability of having a domain of spins $1$
of length $l$ at time $t$.
For $t\to\infty$ $P_t$ converges towards the equilibrium length
probability distribution 
\be
P_{eq}(l)=\frac{\exp(-\beta)}{\bigl [1+\exp(-\beta)\bigr ]^l}
\label{peql}
\ee
and the average length saturates to the equilibrium value
\be
d_{\rm eq}=1+\exp(\beta)
\label{deq}
\ee

\begin{figure}
{\epsfxsize=9cm\epsffile{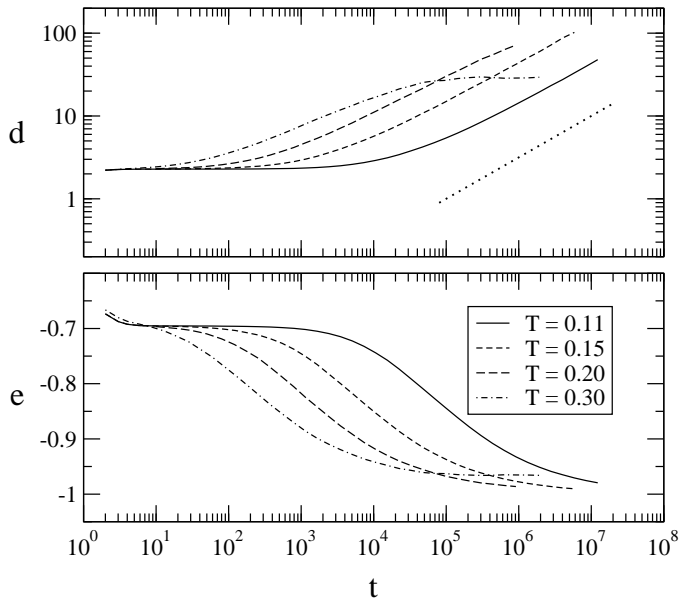}}
\caption{Average domain length and magnetization in the SCIC model.
The average length grows diffusively like $t^{1/2}$.}
\label{d_FA}
\end{figure}

In Fig. \ref{d_FA} we present the average domain length as a function of time
starting from a random initial condition. From the figure it follows that
after $10^6$ MCS $d(t)$ is still well below the equilibrium value 
$d_{\rm eq}$ for temperatures as high as $T=0.2$ indicating that
the systems has not yet equilibrated.
In agreement with \cite{RJ} a
power law fit leads to $d(t)\sim t^{1/2}$ characteristic of 
diffusion \cite{Note1}.
In the lower part of Fig. \ref{d_FA} we show
the relaxation of the energy as a function
of time. These results combine the zero temperature exponential decay
to the threshold energy\cite{FR,ST1} $1/e$  with the slower decay
towards equilibrium. Note that since the average domain length
grows like $t^{1/2}$ while the equilibrium value for large $\beta$
is $d_{\rm eq}\sim \exp(\beta)$ we get  for the
correlation time $\tau_{\rm corr}\sim \exp(2\beta)$, as expected.

\begin{figure}
{\epsfxsize=9cm\epsffile{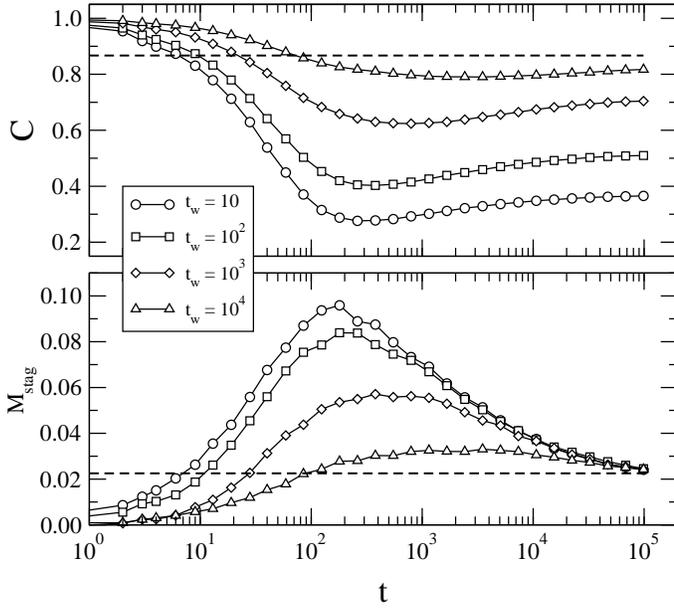}}
\caption{Correlations and zero-field cooled magnetization in a
staggered field in the SCIC model for $N=10^5$, $T=0.3$ and different
waiting times $t_w=10,100,1000,10000$. The horizontal lines indicate the
equilibrium values (\protect\ref{CorrCIC}) and (\protect\ref{MstagCIC}).}
\label{CM_FA_T03}
\end{figure}

\begin{figure}
{\epsfxsize=9cm\epsffile{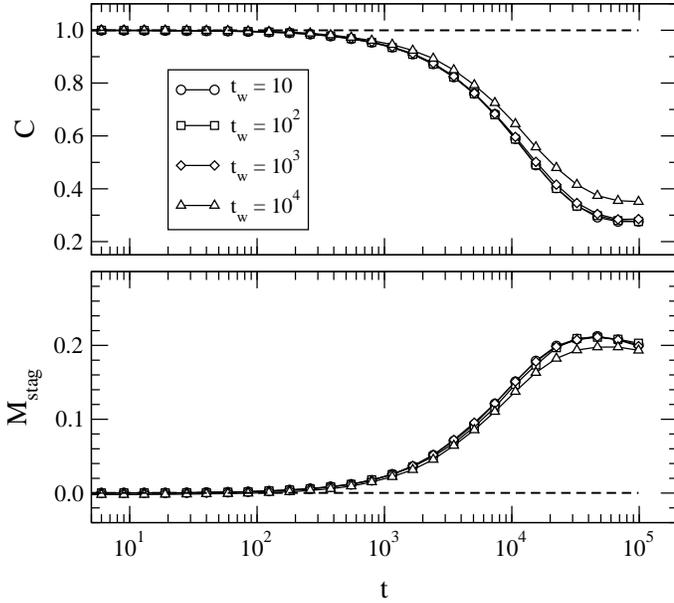}}
\caption{Correlations and zero-field cooled magnetization in a
staggered field in the SCIC model for $N=10^5$, $T=0.11$ and different
waiting times $t_w=10,100,1000,10000$. The horizontal lines indicate the
equilibrium values (\ref{MstagCIC}).}
\label{CM_FA_T011}
\end{figure}

Further informations on the non-equilibrium behavior can be obtained 
from the analysis of the response to a staggered magnetic field as
described in eqs. (\ref{slope}), (\ref{disc_c}) and (\ref{disc_m}).  
In
Figs. \ref{CM_FA_T03} and \ref{CM_FA_T011} we show
$M_{stag}(t+t_w,t_w)$ [see eq. (\ref{disc_m})] and the correlation function
$C(t+t_w,t_w)$ for temperature 
$T=0.3$ and $0.11$ and different waiting times $t_w$. The strength
of the staggered field is $h_0=0.1$ while the system size is $N=10^5$.
The horizontal lines indicate the equilibrium values
\be
C^{eq}=\frac{1}{N}\sum_{i=1}^N\,\langle\nu_i\rangle^2=
\left(\frac{1-\exp(-\beta)}{1+\exp(\beta)}\right)^2
\label{CorrCIC}
\ee
and
\be
M_{stag}^{eq}=\frac{1}{N}\sum_{i=1}^N\,\langle\nu_i\rangle
             =2h_0\frac{\beta\exp(\beta)}
                       {[1+\exp(\beta)]^2}+O(h_0^2)
\label{MstagCIC}
\ee

\begin{figure}
{\epsfxsize=9cm\epsffile{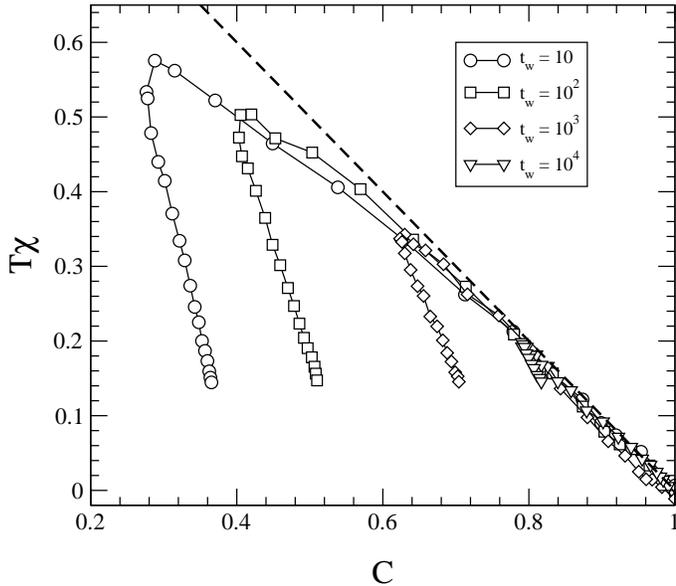}}
\caption{FDT plots in the SCIC for $N=10^5$, $T=0.3$ and different
waiting times $t_w=10,100,1000,10000$. The straight line is the FDT
relation (\ref{eqI2})}
\label{X_FA_T03}
\end{figure}

\begin{figure}
{\epsfxsize=9cm\epsffile{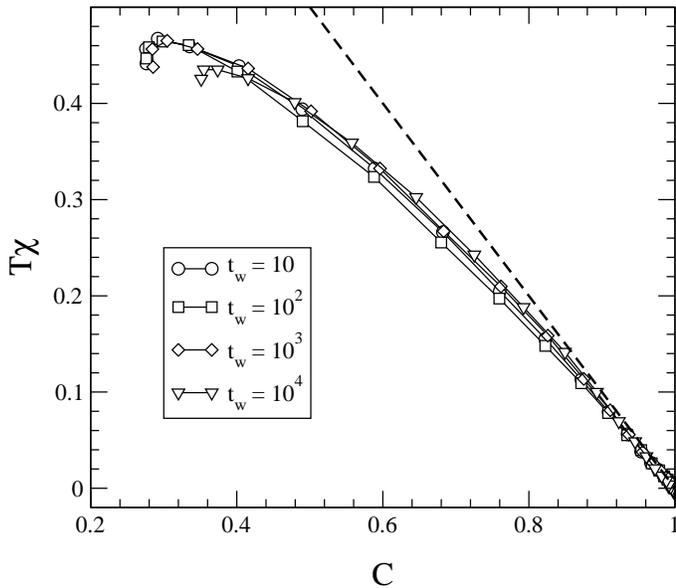}}
\caption{FDT plots in the SCIC for $N=10^5$, $T=0.11$ and different
waiting times $t_w=10,100,1000,10000$. The straight line is the FDT
relation (\ref{eqI2})}
\label{X_FA_T011}
\end{figure}

The corresponding FDT plots are shown
in Figs. \ref{X_FA_T03} and \ref{X_FA_T011}, respectively.

Concerning Figs. \ref{CM_FA_T03}, \ref{CM_FA_T011}, \ref{X_FA_T03}
and \ref{X_FA_T011} we note the following:

\begin{enumerate}
\item Aging in the correlation function appears for 
waiting times larger than the critical
time $\tau_1=\exp{(\beta)}$ and survives even for times larger
than the correlation time $\tau_{\rm corr}$. This can be seen from both
correlation function and staggered magnetization.
The equilibration time $\tau_{\rm eq}$ to reach equilibrium 
is larger than both the activated time $\tau_1$ and the 
correlation time $\tau_{\rm corr}$.
If we define the equilibration time $\tau_{\rm eq}$ as the time
needed for the average domain length to reach the equilibrium
value, then plotting the data of Fig. \ref{d_FA} as function
of $T\log(t/\tau_0)$ we find that 
$\tau_{\rm corr}\sim \frac{\tau_{\rm eq}}{\tau_1}$, i.e.  
$\tau_{\rm eq}\sim \exp(3\beta)$. 

\item The staggered magnetization has a hump, in correspondence of which 
the correlation function presents a broad minimum, 
as function of $t$. For $t_w\le \tau_1$ the hump maximum 
takes the largest value and decreases with $t_w$ as soon as $t_w> \tau_1$
and eventually disappears for $t_w\to \infty$. This effect is
a direct manifestation of the two critical timescales present
in the SCIC model \cite{FR}.

\item The existence of different activated
relaxation times results in rather peculiar FDT plots,
see Figs. \ref{X_FA_T03} and \ref{X_FA_T011}. 
For $t_w<\tau_1$, Fig.  \ref{X_FA_T011}, $C$, $\chi$ and $X$ do not show 
any dependence on $t_w$, nevertheless $X$ is a non-trivial function of $C$
corresponding to non-equilibrium behavior without aging. A similar shape
is found for the one-dimensional Ising
model at low temperatures \cite{ZANETTI}. 
For $t_w > \tau_1$, Fig. \ref{X_FA_T03},  there are
aging effects and $X$ shows the typical two slope pattern. However, 
the existence of
a second typical timescale results in a second downwards bending of the 
IRF and $X$ as function of $C$ has a three slope shape.

\end{enumerate}

\begin{figure}
{\epsfxsize=9cm\epsffile{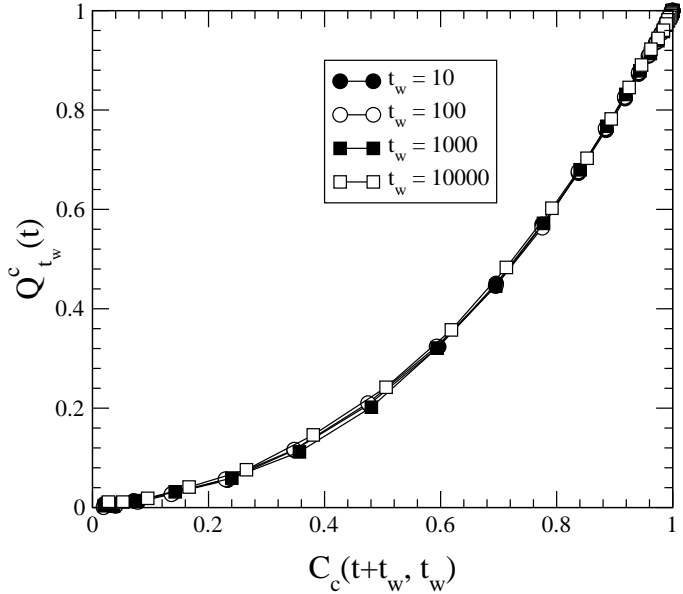}}
\caption{$Q^c_{t_w}(t)$ versus $C_c(t+t_w,t_w)$ in the SCIC at $T=0.11$}
\label{QCFA}
\end{figure}

We conclude the analysis of the SCIC model by discussing the results 
for the cloning experiment.  In Fig. \ref{QCFA} we show the overlap 
$Q^c_{t_w}(t)$ [eq.(\ref{Qcon})] as a function of the connected correlation
$C_c(t+tw,t_w)$ [eq.(\ref{Ccon})] for temperature $T=0.11$. The results
show that, for any finite $t_w$, $Q$ goes to zero quite rapidly. 
If one compares $Q^c_{t_w}(t)$ with $C_c(2t+t_w,t_w)$ for different values
of $t_w$ one finds that both decrease exponentially fast with time and
that $Q^c_{t_w}(t)$ is smaller but very close to $C_c(2t+t_w,t_w)$.
This implies that the two trajectories depart form each other quite fast. 
The data in Fig. \ref{QCFA} collapse quite nicely onto the parabola
$Q\simeq C^2$ in agreement with the exponential decay. 

\begin{figure}
{\epsfxsize=9cm\epsffile{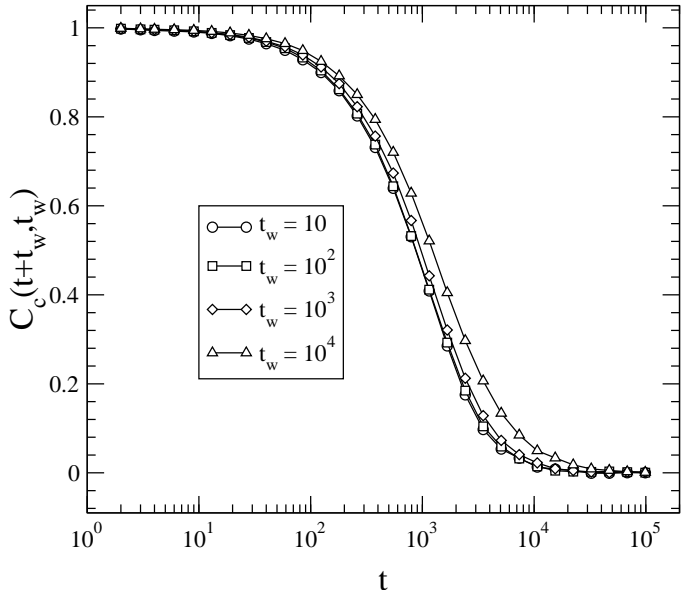}}
\caption{Connected correlations $C_c(t_w,t_w+t)$ in the SCIC for
$N=10^5$, $T=0.15$ and different waiting times
$t_w=10,100,1000,10000$. Aging is present for times larger than
$\tau_1=\exp(\beta)$}
\label{FAcor_t.0.15}
\end{figure}

\begin{figure}
{\epsfxsize=9cm\epsffile{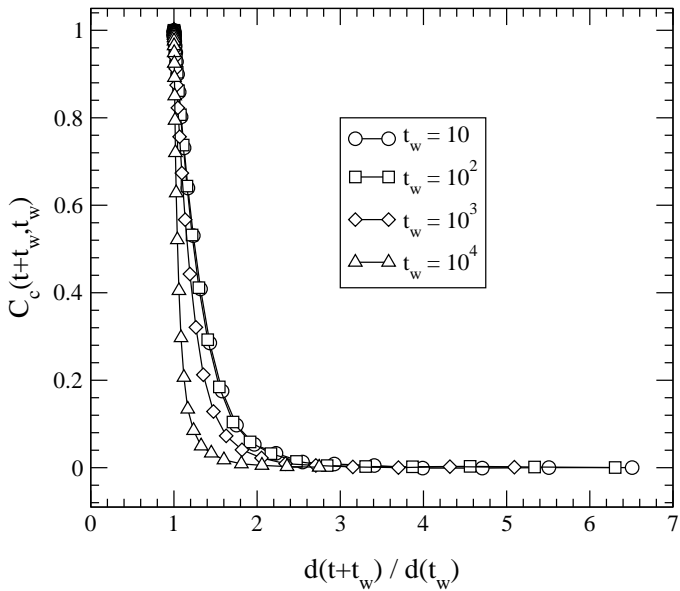}}
\caption{Connected correlations $C_c(t_w,t_w+t)$ (\ref{Ccon}) in the
SCIC for $N=10^5$, $T=0.15$ plotted versus $d(t+t_w)/d(t_w)$.}
\label{FAcor_dttwdtw.0.15}
\end{figure}

In conclusion the SCIC model is a coarsening model with three activated
timescales $\tau_1=\exp(\beta)$, $\tau_{\rm corr}=\exp(2\beta)$ and
$\tau_{\rm eq}=\exp(3\beta)$.  In this scenario we do not expect the
simple scaling form (\ref{eqC}) in terms of a single length scale to be
verified.  Indeed, although a growing length scale can be identified a
single length scale does not describe the whole time regime. This is
clearly seen in Fig. \ref{FAcor_t.0.15} for $T=0.15$ where aging starts
after $\tau_1\sim 1000$. The scaling form (\ref{eqC}) constructed from
data from Fig. \ref{d_FA} is shown in Fig. \ref{FAcor_dttwdtw.0.15}. The
scaling is obviously rather poor.  The behavior of this model is
essentially diffusive as emerges from the behavior of the overlap
$Q^c_{t_w}(t)$ similar to the behavior of a ferromagnet with non
conserved order parameter, with the difference that the SCIC model has
no phase transition and coarsening takes place in a disordered phase.

\subsection{The ACIC model}

Coarsening in this model has been extensively studied by several
authors, finding that the correlation time has a super-Arrhenius
behavior and grows like $\exp(\beta^2/\log(2))$ \cite{MJ,SE}, 
much faster than the typical Arrhenius behavior $\sim\exp(2\beta)$
found for the SCIC model. In the SCIC model domains can
always grow if they can annihilate defects by building intermediate
defects to the right or to the left of that defect. In the ACIC model,
on the contrary, 
defects can disappear only by anchoring intermediate defects in the
middle of the domain from one side. This strongly enhances the
correlation time. On the other hand, the absence of a critical
time like $\tau_1$ and the coincidence of the
correlation time with the equilibration time makes the dynamics of this
model simpler than that of the SCIC model.

\begin{figure}
{\epsfxsize=9cm\epsffile{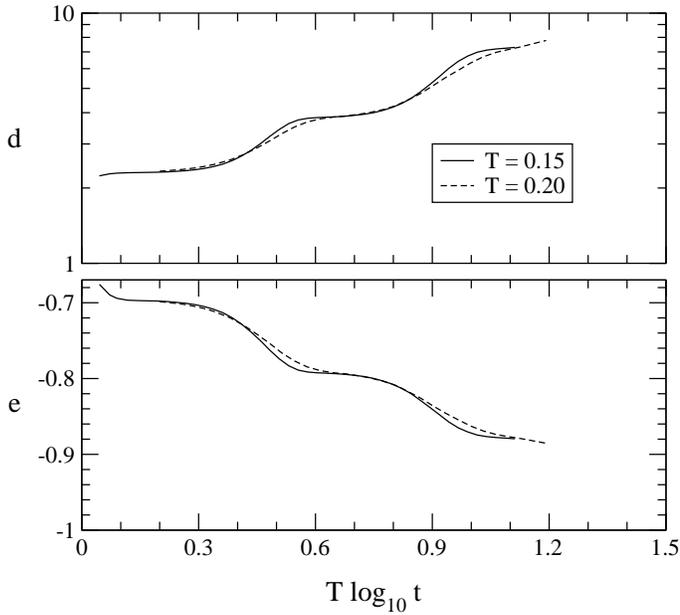}}
\caption{Average domain length and magnetization in the ACIC at
$T=0.15,0.20$. The average length grows like
$t^{\lambda T}$.}
\label{d_SE}
\end{figure}

In Fig. \ref{d_SE} we show the average domain length defined by 
eq. (\ref{d}) and the energy as a function of time 
when starting from a random initial configuration for different
temperatures.  
The $T\log(t)$ scaling  predicted by Sollich and
Evans \cite{SE} is very well satisfied. 
Note the presence of plateaus in both the average
domain length and the energy for the same range of time. 
These correspond to time intervals
where domains coalesce and the global energy stays constant
because the number of anchoring spins is much smaller than the length of
the coalescing domains \cite{SE}. Since the average domain length
grows like $\log(d)\sim \lambda T\log(t)$, i.e.,  $d\sim t^{\lambda T}$,
and the equilibrium domain length is given by $d_{\rm eq}\sim\exp(\beta)$
for low temperature, 
a phenomenological argument yields for the correlation time
$\tau_{\rm corr}\sim \exp(\beta^2/\lambda)$ with
$\lambda=\log(2)$ in agreement with the expectations of Mauch and
J\"ackle \cite{MJ}. 

\begin{figure}
{\epsfxsize=9cm\epsffile{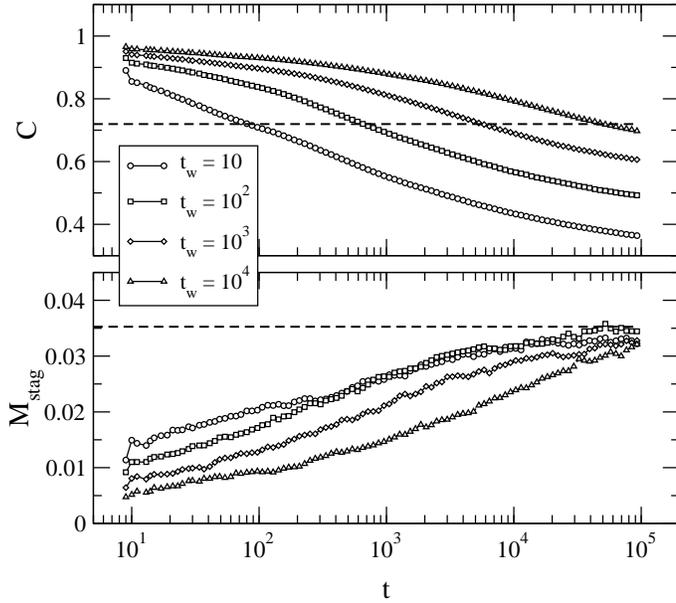}}
\caption{Correlations and zero-field cooled magnetization in a
staggered field in the ACIC model for $N=10^5$, $T=0.4$ and different
waiting times $t_w=10,100,1000,10000$. The horizontal line indicates the equilibrium value
(\ref{MstagCIC})}
\label{M_SE_T04}
\end{figure}

\begin{figure}
{\epsfxsize=9cm\epsffile{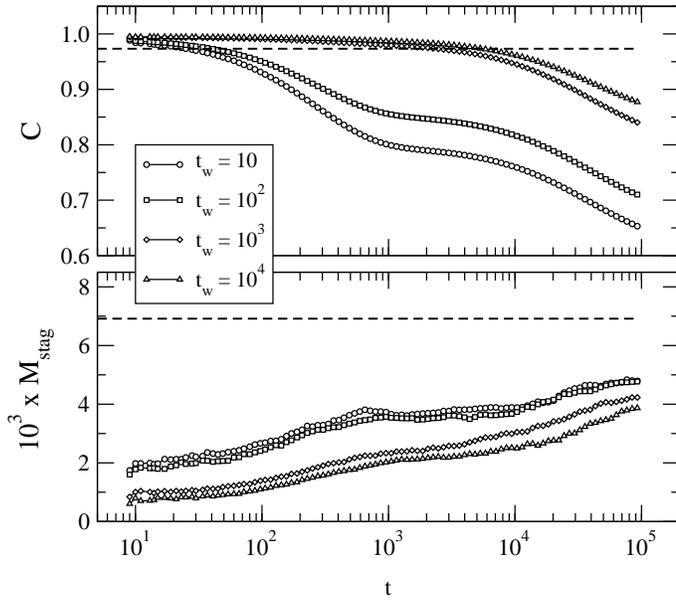}}
\caption{Correlations and zero-field cooled magnetization in a
staggered field in the ACIC model for $N=10^5$, $T=0.2$ and different
waiting times $t_w=10,100,1000,10000$. The horizontal line indicates the equilibrium value
(\ref{MstagCIC})}
\label{M_SE_T02}
\end{figure}

\begin{figure}
{\epsfxsize=9cm\epsffile{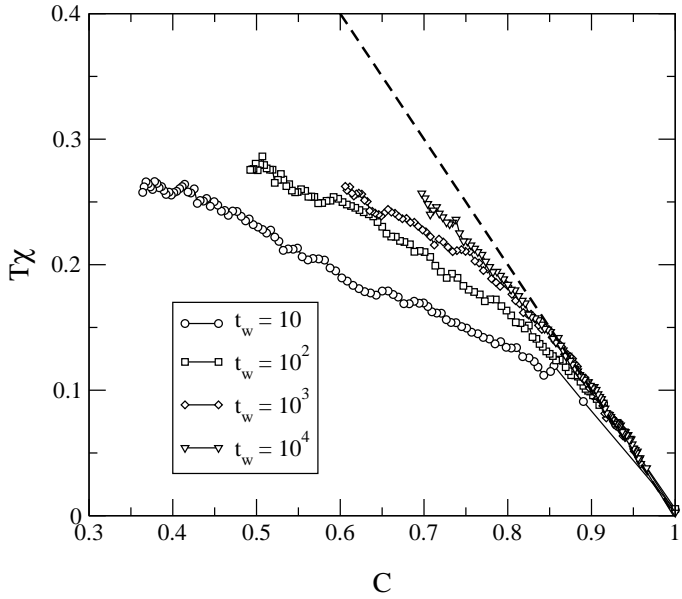}}
\caption{FDT plots in the ACIC for $N=10^5$, $T=0.4$ and different
waiting times $t_w=10,100,1000,10000$. The straight line is the FDT
relation (\ref{eqI2})}
\label{X_SE_T04}
\end{figure}

\begin{figure}
{\epsfxsize=9cm\epsffile{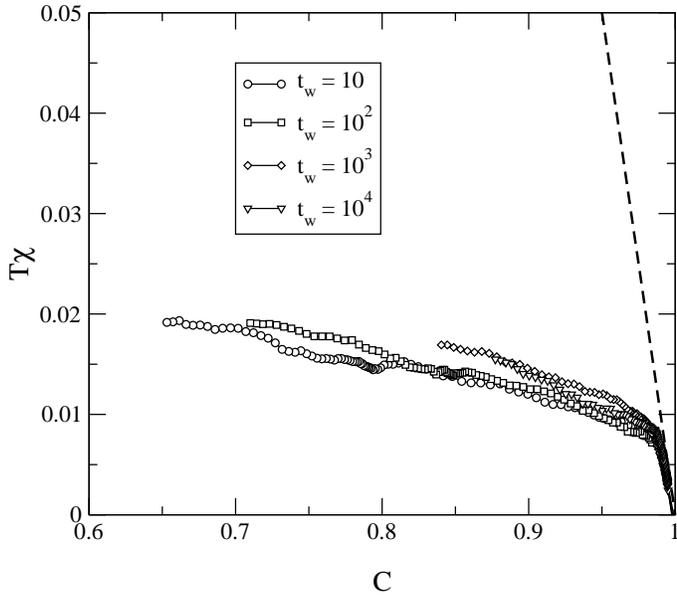}}
\caption{FDT plots in the ACIC for $N=10^5$, $T=0.2$ and different
waiting times $t_w=10,100,1000,10000$. The straight line is the FDT
relation (\ref{eqI2})}
\label{X_SE_T02}
\end{figure}

Figures \ref{M_SE_T04} and \ref{M_SE_T02} show the non-equilibrium 
disconnected correlation function (\ref{disc_c}) and the staggered 
magnetization (\ref{disc_m}) from infinite temperature initial conditions
as a function of time for two temperatures $T=0.2$ and $0.4$ and different 
values of $t_w$. The  strength of the field is $h_0=0.1$.
The dashed horizontal lines are the 
equilibrium values (\ref{CorrCIC}) and (\ref{MstagCIC}). The 
FDT plots are shown in Figs. \ref{X_SE_T04} and \ref{X_SE_T02}.

Looking at this set of figures we note the following points:

\begin{enumerate}

\item The staggered magnetization does not posses a hump, and the correlation
functions a broad minimum, as in the SCIC model. Here, on the contrary,
both are monotonic functions of time, a behavior commonly found in models where
there is no critical time (like $\tau_1$ in the SCIC) associated with a
microscopic fast process. On the timescale of 
$\tau_{\rm corr}$ both quantities relax to the
equilibrium values. We note that
$\tau_{\rm corr}\simeq 10^4$ for  $T=0.4$ and $\tau_{\rm corr}\simeq 10^{15}$
for $T=0.2$.

\item A simple look at Figs. \ref{M_SE_T04} and \ref{M_SE_T02} reveals that
aging is present for all timescales in both the correlation function
and staggered magnetization.  Aging in $M_{stag}$ is noticeable for all 
values of $t_w$ suggesting that aging in the IRF disappears rather slowly with
$t_w$. Keeping in mind the coarsening nature of this model this suggests
that $\alpha\simeq z$ in eq. (\ref{eqI_aging}).

\item From Figs. \ref{M_SE_T04} and \ref{M_SE_T02} it is difficult
to verify the coarsening nature of the dynamics in this model. A simple
check can be done with the help of the FDT plots shown in
Figs. \ref{X_SE_T04} and \ref{X_SE_T02} for  temperatures $T=0.4$
and $T=0.2$.
Interestingly for waiting times
comparable to the correlation time, so that the system is not far from
equilibrium, the fluctuation-dissipation ratio $X$ rapidly converges to $1$,
see Fig. \ref{X_SE_T04}.
At low temperatures, Fig. \ref{X_SE_T02},  $t_w\ll \tau_{\rm corr}$ and the
fluctuation-dissipation ratio is very small, $X\simeq 0.1$, and 
roughly independent of $t_w$, a scenario typical of coarsening
models \cite{BARRAT}.

\end{enumerate}

\begin{figure}
{\epsfxsize=9cm\epsffile{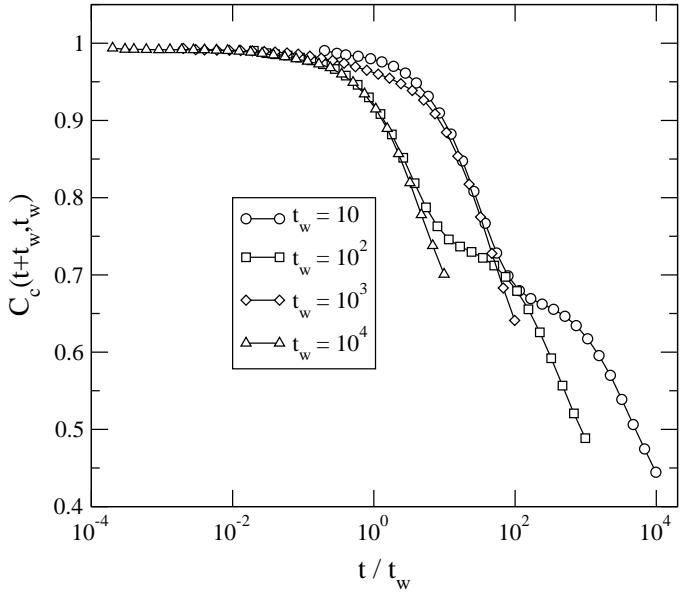}}
\caption{Connected correlations $C_c(t_w,t_w+t)$ in the ACIC for
$N=10^5$, $T=0.20$ and different values of $t_w$ plotted versus $t$}
\label{corr_ttw.0.20}
\end{figure}

\begin{figure}
{\epsfxsize=9cm\epsffile{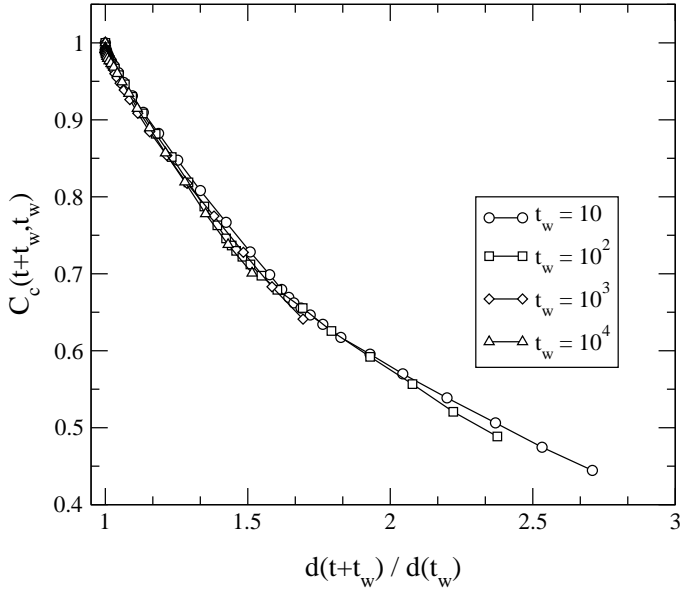}}
\caption{The same data of Fig. \protect\ref{corr_ttw.0.20} plotted versus
$d(t_w+t) / d(t_w)$}
\label{corr_dtdtw.0.20}
\end{figure}

In the ACIC model there is only one characteristic divergent timescale,
namely
$\tau_{\rm corr}$, and thus we expect that the scaling behavior
(\ref{eqC}) should be satisfied. Note that, contrarily to the SCIC, the
average domain length $d(t)$ does not grow in time like a power
law [see Ref. \cite{SE} and Fig. \ref{d_SE}]. 
Consequently, in the aging regime the scaling will not
be of the form $t/t_w$, see Fig. \ref{corr_ttw.0.20},
but a more complicated function
depending on shape of $d(t)$ \cite{SE}, see 
Fig. \ref{corr_dtdtw.0.20} where
$C_c(t_w,t_w+t)$ is plotted as a function of $d(t_w+t)/d(t_w)$. The
scaling is quite good.

\begin{figure}
{\epsfxsize=9cm\epsffile{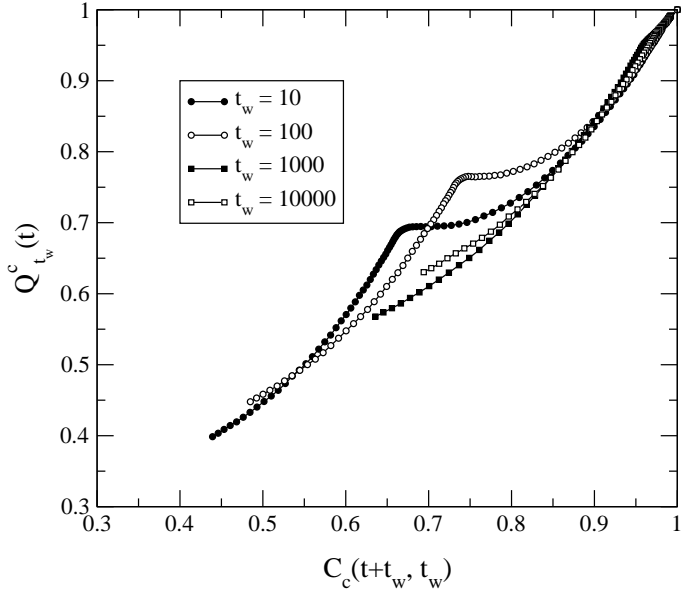}}
\caption{$Q^c_{t_w}(t)$ versus $C_c(t+t_w,t_w)$ in the ACIC at $T=0.2$}
\label{QCSE}
\end{figure}

\begin{figure}
{\epsfxsize=12cm\epsffile{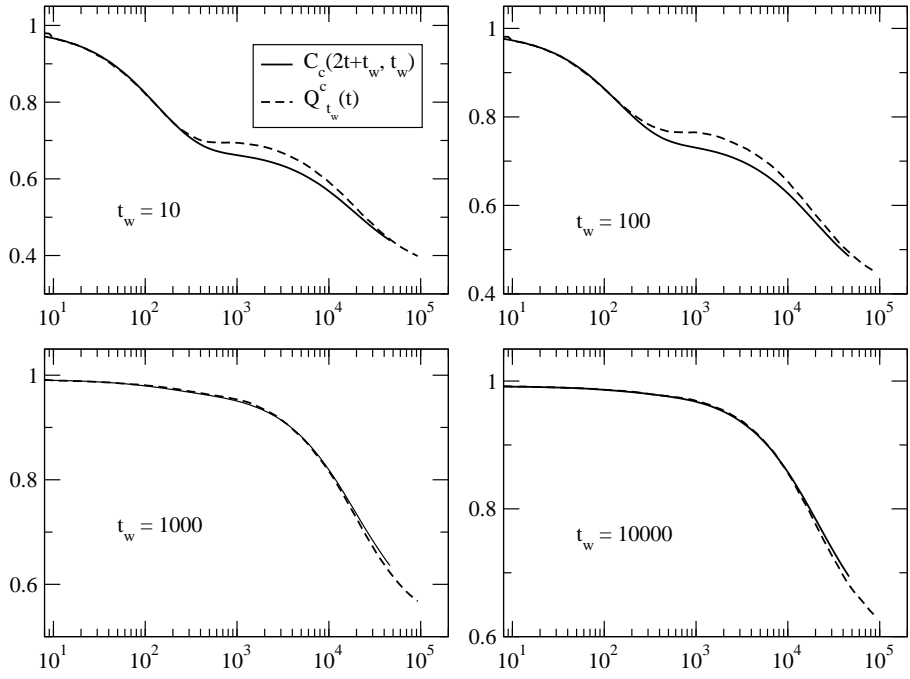}}
\caption{$Q^c_{t_w}(t)$ and  $C_c(2t+t_w,t_w)$ in the ACIC at $T=0.2$}
\label{QCSE2}
\end{figure}

We conclude our analysis of the ACIC model with the discussion 
of $Q^c_{t_w}(t)$ and $C_c(t+t_w,t_w)$ shown in 
Figs. \ref{QCSE} and \ref{QCSE2}. Again 
$Q^c_{t_w}(t)$ decays to zero for any $t_w$ but slower than in the
SCIC model, see Fig. \ref{QCFA}.
If $Q^c_{t_w}(t)$ is compared with $C_c(2t+t_w,t_w)$ for
different values of $t_w$ one finds that again $Q^c_{t_w}(t)$ is smaller,
but very close to, $C_c(2t+t_w,t_w)$. 
However, contrarily to the SCIC case, now during the initial regime 
when $t_w$ is small both $Q^c_{t_w}(t)$ and $C_c(2t+t_w,t_w)$ show a plateau, 
more pronounced for $Q$, for times $t\simeq 10^3$.
The time range  between
$t_w=10^2$ and $t_w=10^3$ corresponds [see Fig. \ref{d_SE}] 
to the regime where $d(t)$ is growing very fast. This
means that during this time interval domains grow, since $C$ 
slowly decays, but $Q$ remains 
almost constant because the two replicas follow the same narrow path 
in phase space. This effect is consequence of the way
domains grow in this model where the anchoring of spins proceeds one by
one in a given direction, different from the diffusive mechanism in
the SCIC model. For $t_w$ larger than $10^3$ this effect would be 
observed at the next timescale, between $10^5$ and $10^6$, where new
domains would have grown again [see Fig. \ref{d_SE}]. 
Moreover this would also lead to 
a new
plateau for $Q$ and $C$ for waiting times 
$t_w=10$ and  $10^2$ for values of the correlation of order $0.3$  
(which we have not reached in the simulations).

We can summarize the results of this subsection by saying that the 
non-equilibrium behavior of the ACIC
model resembles coarsening in a simple ferromagnet
although the growing length scale grows slower,  like
$t^{T\log(2)}$. The ansatz (\ref{eqC}) for the scaling behavior is well
verified and the FDT plots show some similarities with the physics of
undercooled liquids although no connection between the slope of the FDT
behavior and the SW configurational entropy is possible. Note that we
obtain $X\simeq 0$ when temperature is lowered. Furthermore, the
behavior of $Q^c_{t_w}(t)$ reveals that the relaxational dynamics
proceeds by evolution in narrow channels in the time intervals where
domains grow similar to what happens for type I models. 
On the contrary, it is not possible to establish the glassy
scenario from the value of $\lim_{t\to\infty} Q^c_{t_w}(t)$ a quantity
which we expect to work only for models with broken ergodicity.

\subsection{The BG model}

In the 1D BG model, at difference with the kinetic constraint models
discussed above, there are no dynamical constraints
and coarsening follows from the slow growth of the number of empty
boxes induced by  entropic barriers.
If we denote with $\sigma=0,1$ empty and occupied box, respectively, then
we can consider the same quantities discussed for the kinetic
constraint models, i.e., domain length probability
distribution, average length, correlation and magnetization.
All these quantities converge for large times to 
their equilibrium values derived from the
equilibrium probability distribution
\cite{BG-closed},
\be
P_{\rm eq}(n)=\frac{1}{N}\sum_{r=1}^N\langle\delta_{n_r,n}\rangle
=\frac{z^{n-1}\exp(\beta\delta_{n,0})}{n!\exp(z)}
\label{peq}
\ee
\noindent
which gives the probability that at equilibrium a box 
contains $n$ particles. 
Normalization of the distribution (\ref{peq}) corresponds to
the conservation of the total number of particles, and
reads
\be
\exp(\beta)-1=(z-1)\exp(z).
\label{conserv}
\ee
By using $P_{\rm eq}$ the equilibrium values of correlations, 
response and domain length can be computed. 
For example the average domain length of 
empty boxes  at equilibrium is given by,
\begin{equation}
d_{\rm eq}=\frac{1}{1-P_{\rm eq}(0)}
\label{deqBG}
\end{equation}
which diverges for $\beta\to \infty$ when all particles fill a single
box. 
 From the definition it follows that $P_{\rm eq}(0)$ is equal to minus 
the equilibrium  energy (\ref{eq4}) which, for small $T$, goes as 
$E/N=-1+T+O(T^2)$. We then have
$d_{\rm eq}\sim \beta$ for $\beta\to\infty$.

Similarly for the correlation function we have, 
\begin{equation}
C^{\rm eq}=\frac{1}{N}\sum_{i=1}^N\,\langle\nu_i\rangle^2
          =[2P_{\rm eq}(0)-1]^2
\label{Cst_BG}
\end{equation}
where $\nu = 2\sigma - 1$.

To compute the staggered magnetization we have to use the equilibrium
probability distribution with the extra term (\ref{eqP}) added to
the energy. A straightforward calculation leads to
\begin{equation}
M_{stag}^{\rm eq}=\frac{1}{N}\sum_{i=1}^N\,\langle\nu_i\rangle
                 =\frac{e^{\beta}(1-e^{-z})\,\sinh(\beta h_0)}
                       {z\,(e^{z}-1+e^{\beta}\cosh(\beta h_0))}
\label{Mst_BG}
\end{equation}
where $z$ is now solution of
\begin{equation}
\label{eq:zh}
 \frac{ze^z}{2}\,\left[
       \frac{1}{e^z - 1 + e^{\beta(1+h_0)}} +
       \frac{1}{e^z - 1 + e^{\beta(1-h_0)}} 
                 \right] = 1.
\end{equation}
Note that $M_{stag}^{\rm eq}$ is linear in $h_0$ for small values of $h_0$.

Analyzing in details the dynamics we can distinguish 
two different decay processes: the first one entropically activated
and the second one energetically activated. 
When the system is quenched from high to low $T$ particles 
``evaporate'' from some boxes and accumulate in others. After some time
a situation is reached where boxes with more then one particle 
are separated by empty boxes and a very small number of single 
occupied boxes (defects). 
For $T=0$ these defects disappear and the
energy does not relax to equilibrium. On the contrary 
for $T$ small but finite the number of these defects may be large enough,
its number scaling as $TN$, to serve as nucleation paths between two 
nearby multiple occupied boxes which, by
the usual entropic mechanism, eventually accumulate onto a single box. 
The timescale $\tau_1$ for this process is activated since
defects must be created, but it is smaller than $\exp(\beta)$. 

A second, energetically activated, process  
appears for defects to be anchored between
multiple occupied boxes so that they coalesce in a single multiple occupied 
box.
At low temperature and close to equilibrium the typical number of
particles per occupied box is of order $\beta$ while the distance that
particles must cover by diffusion from one box to a contiguous one is
of order $d_{\rm eq}\sim \beta$. Combining these two behaviors
we obtain for the equilibration time
$\tau_{\rm eq}\sim \beta\exp(\beta)$. 

\begin{figure}
{\epsfxsize=9cm\epsffile{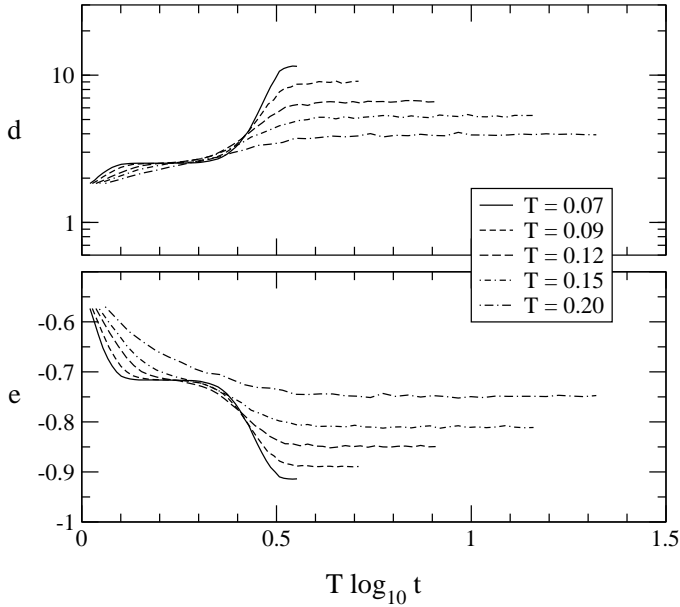}}
\caption{Average domain length and energy in the BG model.}
\label{d_BG}
\end{figure}

The interplay between these two mechanisms can clearly be seen 
in Fig. \ref{d_BG} where we report growth of the average domain length,
from an initial random configuration, and (minus) the average
number of empty boxes (energy) as a function of time for different
temperatures. 
The data are plotted as function of $T\log(t)$ so that the equilibration 
time, where the average length reaches the equilibrium value,
is $T\log(\tau_{\rm eq})\simeq 1+\delta$ with $\delta=-T\log(T)$
leading to $\tau_{\rm eq}\simeq \beta\exp(\beta)$. 
This equilibration timescale $\tau_{\rm eq}$ is shorter than 
that of  the kinetically constrained Ising
chain but longer than the mean-field case 
$\tau_{\rm eq}\sim \frac{\exp(\beta)}{\beta^2}$ \cite{SF}.  From the
figure we also see that both the average domain length and the energy
display a plateau at short times corresponding to the $T=0$
behavior. The departure from this initial plateau occurs for times
shorter than the activated characteristic time $\exp(\beta)$ and 
is driven by the entropic mechanism described above. 
A good collapse of the departing time is obtained for 
$\tau_1=\exp(\beta)/\beta$. For $\beta\to\infty$
the characteristic times $\tau_1$ and $\tau_{\rm eq}$ 
become well separated since 
$\tau_{\rm eq}/\tau_{1}\sim \beta^{2}$.

\begin{figure}
{\epsfxsize=9cm\epsffile{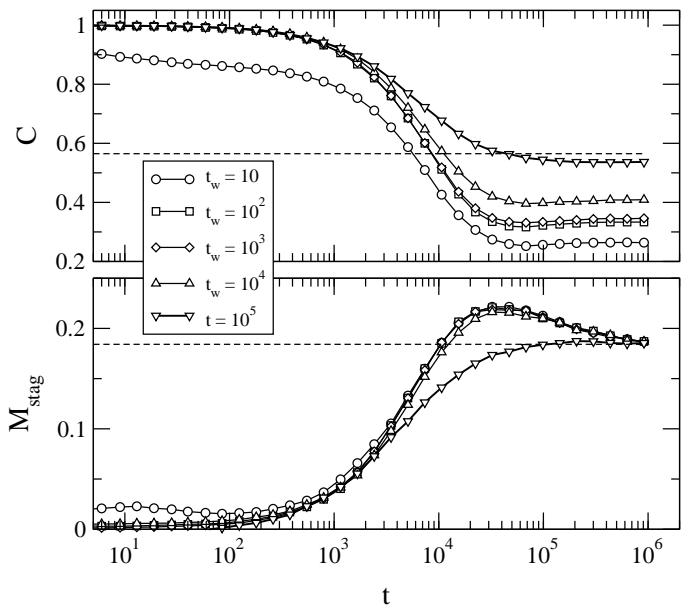}}
\caption{Correlations and staggered magnetization in the BG for
$N=10^4$, $T=0.10$ and different values of $t_w$}
\label{BGsco.0.10}
\end{figure}

\begin{figure}
{\epsfxsize=9cm\epsffile{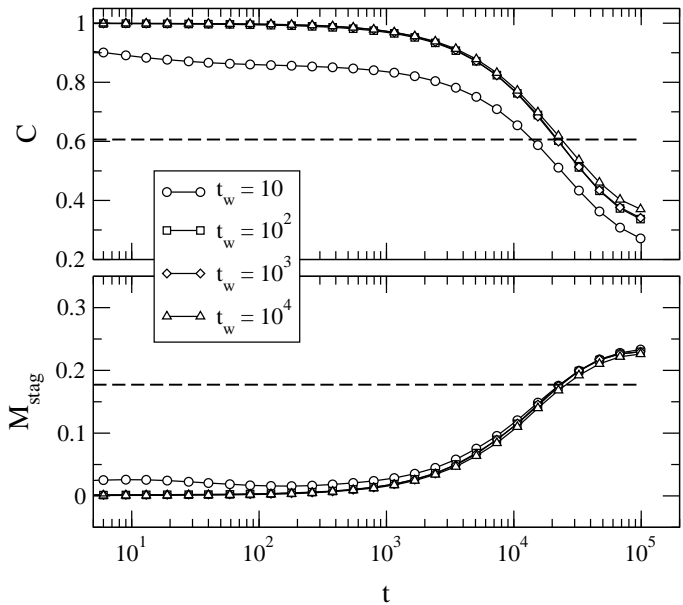}}
\caption{The same as Fig. \protect\ref{BGsco.0.10} for $T=0.09$.}
\label{BGsco.0.09}
\end{figure}

\begin{figure}
{\epsfxsize=9cm\epsffile{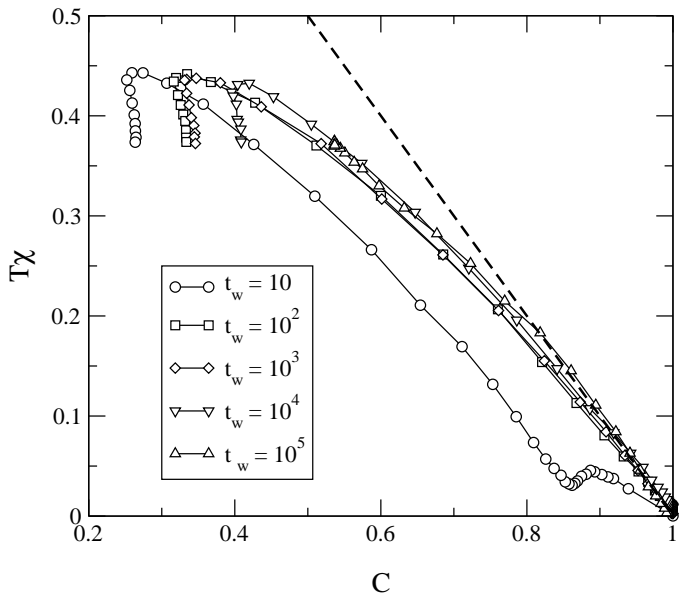}}
\caption{FDT plots in the BG for $N=10^4$, $T=0.1$ and different values
of $t_w$.}
\label{BGfdt.0.10}
\end{figure}

\begin{figure}
{\epsfxsize=9cm\epsffile{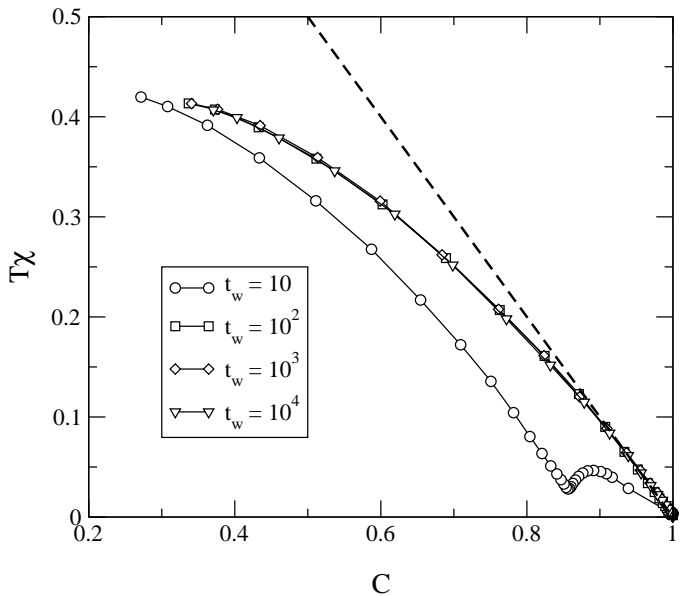}}
\caption{The same as Fig. \protect\ref{BGfdt.0.10} for $T=0.09$.}
\label{BGfdt.0.09}
\end{figure}

Figures \ref{BGsco.0.10}, \ref{BGsco.0.09}, \ref{BGfdt.0.10} and
\ref{BGfdt.0.09} show the correlation function, staggered magnetization and
FDT plots of the 1D BG model for the two temperatures $T=0.1$ and $0.09$ and
field strength $h_0=0.1$.

 From the analysis of the figures the following conclusions can be 
drawn. 

\begin{enumerate}

\item After quenching from infinite to low temperature a 
fast evaporation of occupied boxes occurs after which 
only a finite fraction of  them, approximately $30\%$, survives.
Each occupied box contains in average about $3-4$ particles. 
This process is clearly seen in Fig. \ref{d_BG}.
The timescale  $\tau_1\simeq \exp(\beta)/\beta$ has a role 
similar to the timescale $\tau_1$ found in the SCIC model.
The aging effects are absent for $t_w<\tau_1$ but $X<1$  
[see Figs. \ref{BGfdt.0.10} and \ref{BGfdt.0.09}]. 
The dynamics is again diffusive and similar to the one
found for the one-dimensional Ising model \cite{ZANETTI}.

\item For waiting times $t_w> \tau_{1}$ the dynamics slows down due
to energy barriers. To empty a box a
large number of particles must be transferred, a
process which is cooperative and involves all the particles in
the box. The typical time of this cooperative process 
is $\tau_{\rm eq}\sim \beta\exp(\beta)$.
For waiting times $\tau_1<t_w< \tau_{eq}$ the system
shows strong non-equilibrium effects
with a downwards bending of the IRF as a function $C$ similar to
what seen in the SCIC model. 
The origin of this effect is, however, different 
and follows from the asymmetric response to the staggered field of 
occupied and empty boxes. Since the field is coupled to empty boxes,
the typical time to empty a box is larger than that 
to occupy an empty one.
In other words, when quenching from high (or infinite) temperature
boxes are occupied fast and its number converges relatively fast towards
the equilibrium value. However, due to the staggered field,
the distance among them is far from the equilibrium value and 
occupied boxes must be rearranged, which is a very slow process.
Consequently correlation functions and staggered magnetizations show
peculiar humps corresponding to the fast and slow responses.

\end{enumerate}

In conclusion we can say that the strong entropic effects
which follows from the large occupation numbers of some boxes at low 
temperatures, imply that the non-equilibrium behavior of
the 1D BG model cannot be well 
described only
in terms of a single length scale $d(t)$. The reason is that
although $d(t)$ tells
how far the system is from equilibrium it does not contains enough
informations to efficiently 
describe the effects associated with the entropic barriers and the
ansatz (\ref{eqC}) does not hold.

\section{Conclusions}

In this paper we have studied the dynamics of constrained 1D Ising
models. In particular we have focused on a family of
constrained kinetic models which interpolate between the symmetrically
constrained Ising chain (SCIC) introduced by Fredrickson and Andersen
\cite{FA} and the asymmetrically constrained Ising chain (ACIC)
introduced by Eisinger and J\"ackle \cite{JE}. 
For comparison we have also studied the 1D Backgammon model\cite{BG}
where dynamics is slowed down by the global constraint imposed by the
conservation of the particle number.
Although these models reproduce some generic features of undercooled liquids 
there are, however, important differences.  

First of all since the class of kinetic constrained models defined
in Section II, [see eqs. (\ref{eq1}) and (\ref{eq2})]
have the same Stillinger-Weber configurational entropy [see Section III]
but rather different non-equilibrium behaviors [see Section IV],
we conclude that the IS approach is not appropriate for these
models.
The reason of this failure can be easily understood.  Dynamics in
coarsening systems is usually described in terms of a growing length
scale which measures how close the system is to equilibrium.  
In this scenario configurations at different times
are obviously overlapping since domains at the early times are
contained in the larger domains at later times.  
Consequently the (slow) approach to equilibrium can be described 
within a geometrical picture in terms of the value of the average 
domain length.

In undercooled liquids such a length scale probably does not exist and no
coarsening takes place in the metastable region.  
Here the slowing down of the dynamics follows from an activated dynamics
in a complex energy landscape composed by many valleys \cite{Stillo}.
The SW idea of mapping run-time configurations onto
local minima of energy surface (IS) is to provide a statistical 
description of the valleys.  
The fact that these valleys are uncorrelated is at the basis of 
the potentiality of the approach, a result 
shared by disordered 
mean-field models for the glass transition such as $p$-spin models 
\cite{CGP} or the random energy model \cite{DERRIDA}.  
Obviously coarsening
plays a role for crystallization processes but we know that the
anomalies found in the undercooled regime are also found in disordered
models with a crystal state \cite{BMe,MPR}, so that the role of the crystal
state in the undercooled dynamics can be ignored.  

We have critically discussed the SW configurational entropy and 
its meaning.  A conclusion which is difficult to escape from is that 
any sort of configurational entropy, for example adapted from 
mean-field theory \cite{MG,CMPV}, will have meaning only from a 
dynamical point of view. Efforts in this direction are
the Stillinger and Weber approach itself, the work of 
Nieuwenhuizen \cite{NIEUWENH}, the mean-field
scenario by Franz and Virasoso \cite{FV} and the very recent approach 
proposed by  Biroli and Kurchan \cite{BKn}.

The description of non-equilibrium dynamics in terms of configurational
entropy is valid, in general, if
relaxation proceeds via activated jumps between
uncorrelated configurations not described by any characteristic length
scale. Another possible rephrasing of this conclusion, following the
definitions in \cite{BBM}, is to say that models of type I cannot be
described in terms of any {\em mean-field} like configurational entropy
such as the SW approach. Inherent structures and their statistical
treatment are only useful for models of type II, category to which
structural glasses belong. But as we have shown here, one must be
careful in using this classification for models without phase transition
but with glassy dynamics when ergodicity is not broken like in ordered
(or disordered) ferromagnets. A simple classification based
on the infinite time limit of the $Q_{t_w}(t)$ seems to be well
posed for coarsening ferromagnet-like systems, but for other
slow dynamics it may be inappropriate, see some results in Ref.
\cite{AFR}. Still a careful
examination of the whole dynamical behavior of $Q_{t_w}(t)$, and not
only its infinite time limit, may provide useful information to
evidentiate how coarsening takes place. 

Complementary to this investigation we also made a detailed study of the
coarsening behavior of the 1D constrained kinetic models
and the 1D Backgammon model analyzing their different non-equilibrium 
behaviors.  Such study confirms theoretical 
results already present in the literature, but also 
contains new ones 
concerning the behavior of the average domain length,
in particular for the 1D Backgammon model for which nothing is known,  
of the magnetization, responses and the violation
of fluctuation-dissipation relation. Concerning
correlation functions we have found that in all models studied here 
the $t/t_w$ scaling behavior characteristic of mean-field models is not
observed. However, when there is a single typical time or length scale,
as in the ACIC model,  
the general scaling (\ref{eqC}) and (\ref{eqR}) 
\cite{BRAY} is found.
For the SCIC model we have found that the predicted critical timescale
$\tau_1$ \cite{FR} marks the violation of the scaling behavior
(\ref{eqC}). This peculiar behavior leads to a hump and a broad minimum in the
integrated response function and the correlation respectively, 
leading to a rather unusual fluctuation-dissipation plots. 
This scenario remains valid for values of $a$ not too far from $1/2$, the region
of $a$ close to $1$ (or $0$) is under investigation and will be reported
elsewhere.
A similar
conclusion is valid for the 1D Backgammon model where the presence of
entropic barriers induces a second characteristic timescale. 
In this respect the model with the simpler non-equilibrium dynamics
is the ACIC which presents a
single timescale and the scaling behavior (\ref{eqC}) is well satisfied.
This results suggest that the FDT plots to distinguish one
glassy scenario from another one must be used with caution.
The simple models studied here show interesting FDT plots but
no relevant informations can be obtained 
to interpret the off-equilibrium scenario.
Quite probably the only cases where
quantities such as configurational entropy or FDT ratio are interesting
and meaningful are those where some kind of universality is
expected. To this class belong structural glasses. For this
case, the Stillinger-Weber decomposition provides a statistical and
useful description of some, otherwise inapplicable, mean-field concepts.

In the present research we have tried to clarify the limit of validity
of the SW approach to the description of the glassy dynamics.  The
conclusions that can be drawn from the present study of kinetic
constrained models and the Backgammon model can be generalized and we
expect that the SW approach will fail to describe any coarsening model.
In coarsening models, the information is always contained in a growing
domain length which specifies the stage of evolution of microscopic
configurations. In undercooled liquids where the simplest coarsening
description fails, the SW description is a good alternative which provides
the natural link with mean-field ideas.

\acknowledgements We acknowledge P. Sollich for useful discussion.
(F.R.) has been supported by the Spanish Government
through project PB97-0971.  MS is supported by the European Commission
(contract ERBFMBICT983561).

Note added: After completion of this work we have known of a work \cite{GN}
by J. P. Garrahan and M. E. J. Newman on a three-spin interactions
model on a triangular lattice who find results similar to those
presented here.

\section*{Appendix A: Closure of dynamical equations}

For the constrained kinetic Ising chain the single-spin
zero-temperature dynamics s:

\begin{eqnarray} 
\dot{\tau}_i &=& - \tau_i  (a \tau_{i+1} + b \tau_{i-1})
\end{eqnarray}
with $a,\, b$ such that  $a + b = 1$.
For $k \ge 0$ we get
\begin{eqnarray} 
\frac{d}{dt}(\tau_i \cdots \tau_{i+k})
 &=&
       - a \left(  k \,\tau_i \cdots \tau_{i+k}
        + \tau_{i-1} \cdots \tau_{i+k} \right)  \\
 & &
       - b \left(  k \,\tau_i \cdots \tau_{i+k}
        + \tau_i \cdots \tau_{i+k+1} \right)
\end{eqnarray}
Following~\cite{FR} we define the set of correlations,
\be
C_k(t)=\frac{1}{N}\,\sum_{i=1}^N\tau_i(t)\tau_{i+1}(t)...\tau_{i+k}(t)
\label{eqCC}
\ee
It is simple to see that, for any  $a,\,b$ provided that $a + b = 1$,
the $C_k$ satisfy the equations
\begin{eqnarray} 
        \frac{d C_k}{dt}
 &=&     - k \, C_k(t) - C_{k+1}(t)
\end{eqnarray}
which are identical to those of FA model computed in~\cite{FR}.
The generating function 
$$
G(x,t)=\sum_{k=0}^{\infty}\frac{x^k}{k!}C_k(t)
$$
closes the hierarchy yielding the result~\cite{FR},
\be
G(x,t)=G_0((1+x)\exp(-t)-1)\label{eqG}
\ee
with the initial condition $G_0(x)=G(x,t=0)$. 
In the $t\to\infty$ limit (\ref{eqG}) yields 
$C_k(\infty)=C_0(\infty)\delta_{k,0}$ where,

\be
C_0(\infty)=\sum_{k=0}^{\infty}\frac{(-1)^k}{k!}C_k(0)~~~~~~~.
\label{c0infty}
\ee

\section*{Appendix B: Analytical calculation of $P_{IS}(e,T)$}

Here we give an analytical derivation of the SW configurational entropy
for the constrained Ising chain from the zero-temperature dynamics
described in Appendix A.
If at $t=0$ the system is in thermal equilibrium at temperature $T$
and the system is large enough the central limit
theorem says that the $C_k(0)$ are Gaussian distributed
variables. Consequently, $C_0(\infty)$ is a Gaussian distributed
variable with the first two moments given by,

\be
\langle C_0(\infty)\rangle = \sum_{k=0}^{\infty}\frac{(-1)^k}{k!}\langle
C_k(0)\rangle\label{first_mom}
\ee

\be
\langle C_0^2(\infty)\rangle_c =\langle C_0^2(\infty)\rangle-\langle
C_0(\infty)\rangle^2= \sum_{k=0}^{\infty}\frac{(-1)^{k+l}}{k!l!}\langle
C_k(0)C_l(0)\rangle_c\label{second_mom} \ee

where the subindex $c$ stands for connected correlations. If $m$ stands
for the magnetization $m= 1/[1+\exp(-\beta)]$ then
$C_k(0)=(1-m)^{k+1}$ for a thermalized initial condition at temperature
$T$. This automatically yields, using (\ref{first_mom}) the average IS
energy $\langle e_{IS}\rangle=\langle C_0(\infty)\rangle-1$ yielding,

\be
\langle e_{IS}\rangle=(1-m)\exp(m-1)-1~~~~.
\label{av_e}
\ee

The computation of the second moment (\ref{second_mom}) is also
straightforward and requires the calculation of $\langle
C_kC_l\rangle_c$. A simple calculation gives,

\be \langle C_kC_l\rangle_c=\frac{1}{N}\Bigl (
(k-l-1+\frac{2}{m})(1-m)^{k+1}-(l+k+1+\frac{2}{m})(1-m)^{k+l+2}\Bigr )~~~~.
\label{ckcl}\ee

This expression is valid for $k\ge l$ and $k+l+1\le N$.
After some calculations we finally obtain,

\bea \langle C_0^2(\infty)\rangle_c=\frac{1}{N}\Bigl (
(1-m)^3\exp(m-1)-\frac{4(1-m)^2}{m}\exp(m-1)+\frac{(2-m)(1-m)}{m}I_0(2(1-m))
\nonumber\\
+\,2(1-m)^3\exp(2m-2)-(\frac{m+2}{m})(1-m)^2\exp(2m-2)\Bigr)~~~.
\label{final}
\eea

where $I_0(x)$ is the zeroth order modified Bessel function
$I_0(x)=\sum_{k=0}^{\infty}\,\frac{x^{2k}}{2^{2k}(k!)^2}$. This finally
yields for the IS probability distribution (\ref{eqpe}):

\be
P_{IS}(e,T)=\frac{1}{\sqrt{2\pi\langle C_0^2(\infty)\rangle_c}}\exp\Bigl
( -\frac{(e-\langle e_{IS}\rangle)^2}{2\langle
C_0^2(\infty)\rangle_c}\Bigr )~~~~~~.
\label{eqBPDEa}
\ee
From
eqs. (\ref{av_e},\ref{final}) we can directly obtain the configurational
entropy in two different ways. One is obtained by exact integration of
the IS energy as function of temperature,

\be
s_c(e)=\int_{0}^T\frac{de(T)}{dT}\frac{dT}{T}
\label{sc1a}
\ee

where $e(T)$ is given by the expression
(\ref{av_e}). The other is obtained by integrating the fluctuations,

\be
s_c(e)=\int_{0}^T\frac{\langle C_0^2(\infty)\rangle_c}{T^3}\,dT\label{sc2}
\ee

According to (\ref{eqpe}) both expressions should coincide when
the term $f(\beta,e)$ in (\ref{eqpe})
is independent of the energy $e$. This is generally not true and 
expressions (\ref{sc1}), (\ref{sc2}) are different. 
The interesting remark is
that, in the limit $T\to 0$ both expressions (\ref{sc1}), (\ref{sc2})
and the fix-point approximation (\ref{eqs2}) coincide to first order
in $T$. The reason is that the SW
configurational entropy has full meaning in the limit where
$s_c$ goes to zero and the IS-basins are very narrow containing
few configurations.

\end{document}